\pdfoutput=1
\documentclass{ar2e}
\usepackage{epsfig}
\usepackage{bm}
\usepackage[intlimits,centertags]{amsmath}

\usepackage{amssymb,amsfonts}
\usepackage{gensymb}
\usepackage{mathrsfs}

\newcommand{\bwide}{\begin{widetext}}
\newcommand{\ewide}{\end{widetext}}
\newcommand{\beq}[1]{\begin{equation} \label{(#1)}}
\newcommand{\eeq}{\end{equation}}
\newcommand{\ba}[1]{\begin{eqnarray} \label{(#1)}}
\newcommand{\ea}{\end{eqnarray}}

\begin{document}
\input epsf.tex    


\setlength{\baselineskip}{14pt}
\renewcommand\baselinestretch{1}

\setlength{\oddsidemargin}{0.5in}
\setlength{\evensidemargin}{0.3in}

\jname{Annual Review of Nuclear and Particle Science}
\jyear{(2009)}
\jvol{60}
\ARinfo{ }

\title{IN SEARCH OF EXTRATERRESTRIAL HIGH ENERGY NEUTRINOS}

\markboth{Anchordoqui \& Montaruli}{In Search of Extraterrestrial High Energy Neutrinos}

\author{Luis A. Anchordoqui \affiliation{Department of Physics,\\ University of Wisconsin-Milwaukee, Milwaukee WI 53201, USA} Teresa Montaruli 
\affiliation{Department of Physics,\\ University of Wisconsin-Madison, Madison WI 53706, USA}}


\begin{abstract}
  In this paper we review the search for astrophysical neutrinos.  We
  begin by summarizing the various theoretical predictions which
  correlate the expected neutrino flux with data from other
  messengers, specifically gammas and ultra-high energy cosmic rays.
  We then review the status and results of neutrino telescopes
  in operation and decommissioned, the methods used for data analysis
  and background discrimination. Particular attention is devoted to
  the challenge enforced by the highly uncertain atmospheric muon and
  neutrino backgrounds in relation to searches of diffuse neutrino
  fluxes.  Next, we examine the impact of existing limits on neutrino fluxes
  on studies of the chemical composition of cosmic rays. After that, we show that not only do neutrinos have the potential to discover astrophysical sources, but the huge statistics of atmospheric muons can be a powerful tool as well. We end by discussing the prospects for indirect
  detection of dark matter with neutrino telescopes.
\end{abstract}

\maketitle

\section{INTRODUCTION}
\label{sec1}

It has long been recognized that high energy protons produced in
cosmic ray accelerators would also generate an observable flux of
cosmic neutrinos, mainly through charged pion production in collisions
with the ambient gas or with radiation
fields~\cite{Gaisser:1994yf}.  Viceversa,
the emission of neutrinos necessarily implies the existence of
relativistic baryons, and consequently the acceleration of cosmic
rays. Upcoming catalogues of cosmic ray and neutrino sources would
thus in principle be identical were it not for the different
properties of the messengers themselves.

Neutrinos can escape from the innermost regions of galaxies where light
and other kinds of electromagnetic radiation are blocked by matter and
hence can be tracers of processes which stay hidden to traditional photon 
astronomy. Furthermore, neutrinos carry with them information about
the site and circumstances of their production.  Undeflected and
unabsorbed they can point back to very distant sources, resolving the
origin of the highest energy cosmic rays and the underlying
acceleration mechanism. Besides, unlike gamma-rays that can be
produced via synchrotron radiation, or inverse Compton scattering,
neutrinos provide incontrovertible evidence for acceleration of
baryonic cosmic rays.

MeV neutrino astronomy has been possible for about
45~years~\cite{Bahcall:1997ha}. Thus far two sources of non-atmospheric
neutrinos have been identified: the Sun~\cite{Cleveland:1998nv} and
SN1987a~\cite{Hirata:1987hu}. During the next decade we will be able
to observe the universe using high energy ``neutrino light.''  This will
doubtless illuminate a wondrous new view of the universe.

Candidate sources of high-energy neutrinos are of both galactic and extra-galactic varieties, the latter being expected to dominate at the highest energies just as for the parent cosmic rays. Energy balance considerations suggest that the diffuse galactic flux of neutrinos predominantly originates in supernova remnants~\cite{Erlykin:2005pz}.  Dense molecular clouds, often found in star forming regions where the supernovae (SNe) explode, are particularly efficient at converting protons into pions that decay into gamma rays and neutrinos~\cite{Anchordoqui:2006pb}.  Other possibilities for galactic sources include microquasars~\cite{Levinson:2001as}, $\gamma$-ray binaries~\cite{Aharonian:2005cx}, and pulsar wind nebulae~\cite{Amato:2003kw}.  By far, the most likely extra-galactic sources are active galactic nuclei (AGN)~\cite{Stecker:1991vm} and gamma-ray bursts (GRBs)~\cite{Waxman:1997ti}. [For reviews see e.g.,~\cite{Bednarek:2004ky} and for neutrino flux estimates see \cite{Kistler:2006hp}]. In addition, the so-called cosmogenic neutrinos are guaranteed to exist, their amount being still highly uncertain. These originate from the decay of charged pions produced in the interactions of protons with the  isotropic photon background~\cite{cosmogenic,Engel:2001hd,Anchordoqui:2007fi}.

As for conventional astronomy, neutrino astronomers observe the
neutrino sky through the atmosphere. This is a curse and a blessing;
the background of neutrinos produced by cosmic rays in interactions
with air nuclei provides a beam essential for calibrating the
detectors and demonstrating the neutrino measurement technique. It
also provides an opportunity to probe neutrino
standard oscillations and those arising from new physics, such as violation
of Lorentz invariance~\cite{GonzalezGarcia:2005xw}. Especially unique
is the energy range of the atmospheric neutrino beam
between $10^4~{\rm GeV} \lesssim E_\nu \lesssim
10^8~{\rm GeV}$, not within reach of accelerators, that may reveal the
existence of neutrinos from prompt decays of charmed
mesons and baryons~\cite{Zas:1992ci}. Over the next decade, a data set of the order of one million atmospheric neutrinos will be collected. The
statistics will be so large that mapping of the Earth's interior will
be possible via neutrino tomography~\cite{GonzalezGarcia:2007gg}.

Though neutrino astrophysics is the central motivation to build
neutrino telescopes, ultrahigh energy cosmic neutrinos are also unique
probes of new physics as their interactions are uncluttered by the
strong and electromagnetic forces and can reach center-of-mass
energies of $\sqrt{s} \sim 245$~TeV~\cite{Anchordoqui:2005is}. Rates
for new physics processes, however, are difficult to test since the
flux of cosmic neutrinos is virtually unknown.  It is possible to
mitigate this by using multiple observables which allow one to
decouple effects of the flux and cross
section~\cite{Anchordoqui:2001cg}.  Large neutrino telescopes may also
open a new perspective for indirect searches of dark matter by
detecting neutrinos produced via annihilation of weakly interactive
massive particles trapped within the Sun, or at the center of the
Earth~\cite{Jungman:1994jr}.

The layout of the paper is as follows: in Sec.~\ref{sec2} we overview
the connection between cosmic ray observations and expected neutrino
fluxes; in Sec.~\ref{sec3} we review the status of neutrino telescope
construction, operation and data analysis; in Sec.~\ref{sec4} we
analyze the upper limits on neutrino fluxes set by the different
experiments to assess their scientific impact; in Sec.~\ref{sec5} we
provide a summary and discuss future prospects.

\section{COSMIC RAY  $\bm{\rightleftharpoons}$ NEUTRINO 
CONNECTION}
\label{sec2}

The cosmic ray flux falls as an approximate power-law in energy,
$J_{\rm CR} (E)\propto E^{-\alpha}$, with $\alpha \simeq 2.7$ from
about 10~GeV up to the `knee' in the spectrum at $E \sim 3 \times
10^{6}$~GeV where it steepens to $\alpha \simeq 3.1$; it then steepens
further to $\alpha \simeq 3.25$ at $E \sim 4 \times10^8$~GeV (the so
called `dip') and flattens back to $\alpha \simeq 2.69$ 
at $E \sim 3 \times10^9$~GeV, the so-called  `ankle'~\cite{Anchordoqui:2002hs}. The spectrum
extends up to at least $E \sim 10^{11}$~GeV with a steep spectrum
($\alpha \simeq 4.2$) beyond $E = 5 \times
10^{10}$~GeV~\cite{Abbasi:2007sv}, in accord with the
prediction by Greisen, Zatsepin and Kuz'min
(GZK)~\cite{Greisen:1966jv}.

The energy density associated with the flux of galactic cosmic rays is
$\rho_E \sim 10^{-12}$~erg cm$^{-3}$~\cite{Gaisser:1990vg}. This is
about the value of the corresponding energy density $B^2/8\pi$ of the
microgauss magnetic field in the Galaxy. Given that the average
containment time of the cosmic rays in our Galaxy is $\sim
3\times10^6$~yr, the power needed to maintain this energy density
is $\sim 10^{-26}$\,erg/cm$^3$s. For a nominal volume of the galactic
disk of $10^{67}$\,cm$^3$ this requires an accelerator delivering
power of about $10^{41}$\,erg/s. This happens to be 10\% of the power
produced by supernovae, releasing $10 ^{51}\,$erg every 30~yr. This 
coincidence is the basis for the idea that shocks produced by
supernovae expanding into the interstellar medium are the origin of
the galactic cosmic rays.
 
The conversion of $10 ^{50}\,$erg of energy into particle
acceleration is believed to occur by diffusive shock acceleration in
the young ($10^3-10^4$~yr) SN remnant expanding into the interstellar
medium. If high energy cosmic rays are indeed associated with the
remnant, they will interact with hydrogen atoms in the interstellar
medium to produce pions that decay into roughly equal numbers of
photons and neutrinos: the former via $\pi^0 \to \gamma \gamma$, the
latter via $\pi^+ \rightarrow e^+ \nu_e \nu_\mu \overline \nu_\mu$
(and the conjugate process). These may provide us with indirect
evidence for cosmic ray acceleration. The observation of these pionic
gamma rays has been one of the motivations for neutrino astronomy as well as, to a large extent,
for gamma-ray astronomy.

Whereas the details are complex and predictions can be treacherous, a
simple estimate of the gamma ray flux associated with a supernova
remnant can be made following Aharonian, Drury and Volk~\cite{Felix}.
The emissivity in pionic gamma rays is simply proportional to the
density of cosmic rays for energy in excess of 1~TeV, $n_{\rm
  CR}(>1~{\rm TeV}) \simeq4\times10^{-14}\, {\rm cm}^{-3},$ and to the
target density of hydrogen atoms, $n_{\rm H}$. The proportionality
factor is determined by particle physics.

The emissivity (number/volume/time/energy) of neutral pions resulting from an
isotropic distribution of highly relativistic particles following a
power-law energy spectrum $dn_{\rm CR}(E_N)/dE_N \propto E_N^{-\alpha}$ is
given by
\begin{eqnarray}
\label{Qpi}
Q_{\pi} (E_{\pi}) & = &  c \, n_{\rm H} \, \int_{E_N^{\rm th}
  (E_{\pi})}^{E_N^{\rm max}} \frac{dn_{\rm CR}}{dE_N} (E_N) 
  \frac{d\sigma_A}{dE_{\pi}} (E_{\pi},E_N) \, dE_N
\end{eqnarray}
where $E_N^{\rm th} (E_{\pi})$ is
the minimum energy per nucleon required to produce a pion with energy
$E_{\pi}$, and $d\sigma_A(E_{\pi},E_N)/dE_{\pi}$ is the
differential cross section for the production of a pion with energy
$E_{\pi}$ in the lab frame due to the collision of a nucleus $A$ of
energy per nucleon $E_N$ with a hydrogen atom at rest. The differential
cross-section can be parametrized by
\begin{equation}
\frac{d\sigma_A}{dE_{\pi}} (E_{\pi},E_N) \simeq
\frac{\sigma_0^A}{E_{\pi}} \, x F_{\pi}(x, E_N) \,,
\label{parametrization}
\end{equation}
where $x \equiv E_{\pi}/E_N$, $\sigma_0^A = A^{2/3} \, \sigma_0$ (with
$\sigma_0 \simeq 40$~mb) provides a scaling of the cross-section with
the atomic mass number~\cite{Gaisser:1990vg}, and $F_{\pi}(x, E_N)$ is
a fragmentation function. Taking into account that inelastic hadronic
collisions lead to roughly equal numbers of $\pi^0$, $\pi^+$, and
$\pi^-$ mesons, as well as the mild variation with pseudorapidity, the
parametrization~\cite{Kelner:2006tc}
\begin{eqnarray}
F_{\pi} (x, E_N) & = & 4 \beta B_\pi x^{\beta -1} \,
  \left(\frac{1-x^{\beta}}{1+ r \, x^{\beta} \, (1-x^{\beta})}\right)^4 
  \left(\frac{1}{1-x^{\beta}} + \frac{r \, (1-2 x^{\beta})}{1 + r \,
  x^{\beta} \, (1 - x^{\beta})}\right) \nonumber \\
& \times & \left(1 - \frac{m_\pi}{xE_N}\right)^{1/2}\,,
\label{fpi}
\end{eqnarray}
is found~\cite{Anchordoqui:2009tya} to be consistent at $1\sigma$
level with data collected at the Tevatron by the CDF
detector~\cite{Abe:1989td}; $B_\pi = a + 0.25$, $\beta =
0.98/\sqrt{a},$ $r= 2.6/\sqrt{a},$ $a = 3.67 + 0.83L + 0.075 L^2,$ and
$L = \ln (E_N/{\rm TeV})$. Substitution of Eq.~(\ref{parametrization})
into Eq.~(\ref{Qpi}) leads to
\begin{equation}
Q_{\pi} (E_{\pi}) \simeq  Z_{A \pi} (\alpha) \, Q_A^{A {\rm H}} (E_\pi)
\end{equation}
where
\begin{equation}
Q_A^{A {\rm H}} (E_N) = \sigma_0^A \, c \, n_{\rm H} \, \frac{dn_{\rm CR}}{dE_N} (E_N)
\end{equation}
and the spectrum-weighted moment of the inclusive cross-section or
so-called $Z$-factor is given by
\begin{equation}
Z_{A \pi} (\alpha) \equiv \int_0^1 x^{\alpha-1} \,  F(x,E_N) \,
dx \, .
\label{spectralmoments}
\end{equation}
On average, the photons carry one-half of
the energy of the pion, and thus the $\gamma$-ray emissivity is
\begin{eqnarray}
Q_{\gamma}^{Ap} (E_\gamma) & = & 2 \, \int_{E_{\pi}^{\rm min}
  (E_\gamma)}^{E_{\pi}^{\rm max} } \, \frac{Q_{\pi}^{Ap}
  (E_{\pi})}{\left(E_{\pi}^2 - m_{\pi}^2 \right)^{1/2}} \,
  dE_{\pi} \nonumber \\
 & \simeq & Z_{\pi \gamma} (\alpha) \,
Q_{\pi}^{Ap} (E_\gamma) \, ,
\label{iiioi}
\end{eqnarray}
where $E_{\pi}^{\rm min} (E_\gamma) = E_\gamma +
m_{\pi}^2/(4E_\gamma)$ and $Z_{\pi \gamma} (\alpha) = 2 / \alpha$.
Assuming proton dominance and emission spectrum $\propto E_N^{-2}$,
the gamma ray emissivity above 1~TeV becomes
\begin{eqnarray}
  Q_\gamma (> 1\,{\rm TeV}) &=& c \, \langle E_\pi/E_N \rangle\,\, 
\sigma_0\,\, n_{\rm H}
\, n_{\rm CR}(>1\,{\rm TeV}) \nonumber \\
 & \simeq & 10^{-29} \,
 \left(\frac{n_{\rm H}}{{\rm cm}^{-3}}\right)\,\,
\frac{{\rm photons}}{{\rm cm}^{3} \, {\rm s}}\,\,,
\end{eqnarray}
where $\langle E_\pi/E_N\rangle \sim 0.2$ is the average energy of the
secondary pions relative to the cosmic ray protons. For different spectral
indices, the quantity $\langle E_\pi/E_N\rangle$ is generalized to the
spectrum-weighted moments for pion production by nucleons given in
Eq.~(\ref{spectralmoments}).

The density of protons from a supernova converting a total kinetic
energy $W$ of $10^{50}$\,erg to proton acceleration is approximately
given by $\rho/W$, where we will assume that the density in the
remnant is not very different from the ambient energy density $\rho_E
\sim 10^{-12}$\,erg\,cm$^{-3}$ of galactic cosmic rays. This
approximation is valid for young remnants in their Sedov
phase~\cite{Torres:2002af}.  The total luminosity in gamma rays is
then
\begin{equation}
{\cal L}_{\gamma} (> 1~{\rm TeV})  = 
Q_\gamma\, {W \over \rho_E} \simeq 10^{33}~{\rm photons}\, {\rm s}^{-1} \,\, .
\end{equation}
The expected rate of TeV photons from a supernova at a distance 
$d \sim 1~{\rm kpc}$ is
%
\begin{eqnarray}
\frac{d{\cal N}_\gamma}{d( \ln E_\gamma)}(>1~{\rm TeV}) &  = &  
{{\cal L}_\gamma \over 4\pi d^2} \nonumber \\
 & \simeq & 10^{-11} \left(\frac{{\rm photons}}{{\rm cm}^2 {\rm s}}\right) 
\left({W \over \rm 10^{50}erg} \right) \left({n_{\rm H}\over {\rm cm}^{-3}}
\right) \left({{\rm kpc}\over d}\right)^{2}.
\label{galactic1}
\end{eqnarray}
Neglecting absorption effects, the all flavor neutrino and  photon
fluxes are roughly equal, and so we anticipate an event rate of 3 detected neutrinos per decade
of energy per km$^2$~yr, a result readily obtained from the relation
\begin{eqnarray}
{d{\cal N}_\nu \over d(\ln E_\nu)} (>1~{\rm TeV}) & \simeq & 10^{-11}\left({{\rm neutrinos}\over {\rm cm}^2 \, {\rm s}}\right) \, \left(\frac{\rm area}{1~{\rm km^2}}\right) \,  \left( \frac{\rm time}{1~{\rm yr}}\right) \, P_\nu \,\,,
\label{galactic2}
\end{eqnarray}
where the last factor indicates the average probability for neutrino
detection; for the TeV energy considered here $P_\nu\sim
10^{-6}$~\cite{Gaisser:1994yf}.

This estimate may be somewhat optimistic because we assumed that the
neutrino energy reaches 100~TeV with an $E_\nu^{-2}$
spectrum.\footnote{Such a spectral index is typical of Fermi
  acceleration mechanisms~\cite{Fermi:1949ee}.} On the other hand, if
the galactic cosmic ray spectrum extends at least up to the knee ($E
\sim 3000$~TeV), then some of the sources must produce $\sim 100$~TeV
secondaries.

Photo-disintegration of high-energy nuclei, followed by prompt
photo-emission from the excited daughter nuclei also produces TeV
gamma rays and antineutrinos. In this chain reaction, the nuclei act
in analogy to Einstein's relativistic moving mirror to
``double-boost'' eV~starlight to TeV energies for a Lorentz boost
factor $> 10^6$~\cite{Anchordoqui:2006pd}. The TeV antineutrino
counterpart is produced through $\beta$-decay of the emitted
neutrons~\cite{Anchordoqui:2003vc}.  Effective nucleus
photo-disintegration is expected in stellar associations harboring a
large population of massive and early type stars that can provide the
required Lyman-$\alpha$ emission. Directional beams of galactic
antineutrinos and gamma rays may thus emerge as fluctuations over the
diffuse intensity.

A plethora of explanations have been proposed to address the
production mechanism of cosmic rays beyond the
dip~\cite{Torres:2004hk}. In the absence of a single model which is
consistent with all data, the origin of these particles remains a
mystery. Clues to solve the mystery are not immediately forthcoming
from the data, particularly since various experiments report mutually
inconsistent results.  In recent years, a somewhat confused picture
regarding the nature of these cosmic rays has been emerging. The HiRes
data have been interpreted as a change in cosmic ray composition, from
heavy nuclei to protons, at $E \sim 4 \times
10^8$~GeV~\cite{Bergman:2004bk}. This is an order of magnitude lower
in energy than the previous crossover deduced from the Fly's Eye
data~\cite{Bird:1993yi}.

The end-point of the galactic flux is expected to be dominated by
iron, as the large charge $Ze$ of heavy nuclei reduces their Larmor
radius (containment scales linearly with $Z$) and facilitates their
acceleration to highest energy (again scaling linearly with $Z$).  The
dominance of nuclei in the high energy region of the galactic flux
carries the implication that any changeover to protons represents the
onset of dominance by an extra-galactic component.  The inference from
this HiRes data is therefore that the extra-galactic flux is beginning
to dominate the galactic flux already at $E \sim 4 \times
10^8$~GeV. Such a transition would appear to require considerable
fine-tuning of the shape and normalization of the two spectra. It has
been argued however that the spectral shape required for the
extra-galactic component can develop naturally during the propagation
of extra-galactic protons in the cosmic microwave background (CMB)
over cosmological distances~\cite{Berezinsky:2002nc}. At $E \sim
3\times10^8$~GeV the energy losses due to $e^+e^-$ pair production and
cosmic expansion are roughly equal and thus produce a steepening of an
initially featureless power-law injection spectrum. Significantly, the
onset of the extra-galactic component would be well below $E_{\rm GZK}
\sim 5 \times 10^{10}$~GeV, the threshold energy for resonant $p
\gamma_{\rm CMB} \rightarrow \Delta^+ \rightarrow N \pi$ energy-loss
on the CMB, and so samples sources even at large redshift.

It is helpful to envision the cosmic ray engines as machines where
protons are accelerated and (possibly) permanently confined by the
magnetic fields of the acceleration region. The production of neutrons
and pions and subsequent decay produces neutrinos, gamma rays, and
cosmic rays. If the neutrino-emitting source also produces high energy
cosmic rays, then pion production must be the principal agent for the
high energy cutoff on the proton spectrum.  Conversely, since the
protons must undergo sufficient acceleration, inelastic pion
production needs to be small below the cutoff energy; consequently,
the plasma must be optically thin. Since the interaction time for
protons is greatly increased over that of neutrons because of magnetic
confinement, the neutrons escape before interacting, and on decay give
rise to the observed cosmic ray flux. The foregoing can be summarized
as three conditions on the characteristic nucleon interaction time
scale $\tau_{\rm int}$; the neutron decay lifetime $\tau_n$; the
characteristic cycle time of confinement $\tau_{\rm cycle}$; and the
total proton confinement time $\tau_{\rm conf}$: $(i)\; \tau_{\rm
  int}\gg \tau_{\rm cycle}$; $(ii)\; \tau_n > \tau_{\rm cycle}$;
$(iii)\; \tau_{\rm int}\ll \tau_{\rm conf}$. The first condition
ensures that the protons attain sufficient energy.  Conditions $(i)$
and $(ii)$ allow the neutrons to escape the source before
decaying. Condition $(iii)$ permits sufficient interaction to produce
neutrons and neutrinos. These three conditions together define an
optically thin source~\cite{Ahlers:2005sn}. A desirable property to
reproduce the almost structureless energy spectrum is that a single
type of source will produce cosmic rays with a smooth spectrum across
a wide range of energy.

The cosmic ray flux above the ankle is often summarized as ``one $3
\times 10^{10}$~GeV particle per kilometer square per year per
steradian.'' This can be translated into an energy
flux~\cite{Gaisser:1997aw}
\begin{eqnarray}
E \left\{ E{J_{\rm CR}} \right\} & = & {3 \times 10^{10}\,{\rm GeV} 
\over \rm (10^{10}\,cm^2)(3\times 10^7\,s) \, sr} \nonumber \\
 & = &  10^{-7}\rm\, GeV\ cm^{-2} \, s^{-1} \, sr^{-1} \,.
\end{eqnarray}
From this we can derive the energy density $\rho_E$ in ultra-high
energy cosmic rays using flux${}={}$velocity${}\times{}$density, or
\begin{equation}
4\pi \int  dE \left\{ E{J_{\rm CR}} \right\} =  c\rho_E\,.
\end{equation}
This leads to
\begin{equation}
  \rho_E = {4\pi\over c} \int_{E_{\rm min}}^{E_{\rm max}} { 10^{-7}\over E} 
  dE \, {\rm {GeV\over cm^3}} \simeq 3 \times 10^{-19} \, 
{\rm {TeV\over cm^3}} \,,
\end{equation}
taking the extreme energies of the accelerator(s) to be $E_{\rm max} /
E_{\rm min} \simeq 10^3$. The power required for a population of
sources to generate this energy density over the Hubble time of
$10^{10}$~yr is
\begin{equation}
\left.E^2 \frac{d\dot{N}_{\rm{CR}}}{dE}\right|_{E_{\rm min}} 
= \frac{\dot \epsilon_{\rm CR}^{[10^{19}, 10^{21}]}}{\ln(10^{21}/10^{19})} 
\approx 10^{44}\,\rm{erg}\,\rm{Mpc}^{-3} \rm{yr}^{-1} \,\,,
\end{equation}
where an energy spectrum $\propto E^{-2}$ has been 
assumed~\cite{Waxman:1995dg}.

This works out to about $3 \times 10^{39}\rm\,erg\ s^{-1}$
per galaxy, $3 \times 10^{42}\rm\,erg\ s^{-1}$ per cluster of
galaxies, $\sim 2 \times 10^{44}\rm\,erg\ s^{-1}$ per active galaxy,
or $\sim 2 \times 10^{52}$\,erg per cosmological gamma ray
burst~\cite{Gaisser:1997aw}. The coincidence between these numbers and
the observed output in electromagnetic energy of these sources
explains why they have emerged as the leading candidates for the
cosmic ray accelerators. The coincidence is consistent with the
relationship between cosmic rays and photons built into the previously 
discussed "optically thin" source. 

The energy density of neutrinos produced through $p\gamma$
interactions of these protons can be directly tied to the injection
rate of cosmic rays
\begin{equation}
E^2_{\nu} \frac{dN_{\nu}}{dE_{\nu}}
\approx \frac{3}{8} \epsilon_\pi \, t_{\rm{H}}\,E^2 
\frac{d\dot{N}_{\rm{CR}}}{dE},
\end{equation}
where $t_{\rm{H}}$ is the Hubble time and $\epsilon_\pi$ is the
fraction of the energy which is injected in protons lost into photo-pion
interactions.  (The factor of 3/8 comes from the fact that, close to
threshold, roughly half the pions produced are neutral, thus not
generating neutrinos, and one quarter of the energy of charged pion
decays goes to electrons rather than neutrinos.)\footnote{The average
  neutrino energy from the direct pion decay is $\langle E_{\nu_\mu}
  \rangle^\pi = (1-r)\,E_\pi/2 \simeq 0.22\,E_\pi$ and that of the
  muon is $\langle E_{\mu} \rangle^\pi = (1+r)\,E_\pi/2 \simeq
  0.78\,E_\pi$, where $r$ is the ratio of muon to the pion mass
  squared. In muon decay, since the $\nu_\mu$ has about 1/3 of its parent energy,
  the average muon neutrino energy is $\langle E_{\nu_\mu} \rangle^\mu =(1+r)E_\pi/6=0.26 \,
  E_\pi$.} The
``Waxman-Bahcall bound'' is defined by the condition $\epsilon_\pi
=1$
\begin{eqnarray} [E^2_{\nu} \Phi_{\nu}]_{\rm WB} & \approx & (3/8)
  \,\xi_Z\, \epsilon_\pi\, t_{\rm{H}}\, \frac{c}{4\pi}\,E^2
  \frac{d\dot{N}_{\rm{CR}}}{dE} \nonumber \\ & \approx & 2.3
  \times 10^{-8}\,\epsilon_\pi\,\xi_Z\, \rm{GeV}\,
  \rm{cm}^{-2}\,\rm{s}^{-1}\,\rm{sr}^{-1},
\label{wbproton}
\end{eqnarray}
where the parameter $\xi_Z$ accounts for the effects of source
evolution with redshift, and is expected to be $\sim
3$~\cite{Waxman:1998yy}. For interactions with the ambient gas (i.e.,
$pp$ rather than $p \gamma$ collisions), the average fraction of the
total pion energy carried by charged pions is about $2/3$, compared to
$1/2$ in the photo-pion channel. In this case, the upper bound given
in Eq.~(\ref{wbproton}) is enhanced by 33\%~\cite{Anchordoqui:2004eb}.

The actual value of the neutrino flux depends on what fraction of the
proton energy is converted to charged pions (which then decay to
neutrinos). To quantify this, we follow Waxman-Bahcall and define
$\epsilon_\pi$ as the ratio of charged pion energy to the {\em
  emerging} nucleon energy at the source.  For resonant
photoproduction, the inelasticity is kinematically determined by
requiring equal boosts for the decay products of the
$\Delta^+$~\cite{Stecker:1968uc}, giving $\epsilon_\pi = E_{\pi^+}/E_n
\approx 0.28$, where $E_{\pi^+}$ and $E_n$ are the emerging charged pion
and neutron energies, respectively.  For $pp\rightarrow NN + {\rm
  pions},$ where $N$ indicates a final state nucleon, the inelasticity
is $\approx 0.6$~\cite{Alvarez-Muniz:2002ne}. This then implies that
the energy carried away by charged pions is about equal to the
emerging nucleon energy, yielding (with our definition)
$\epsilon_\pi\approx 1.$

At production, if all muons decay, the neutrino flux consists of equal fractions of
$\nu_e$, $\nu_{\mu}$ and $\bar{\nu}_{\mu}$. Originally, the
Waxman-Bahcall bound was presented for the sum of $\nu_{\mu}$ and
$\bar{\nu}_{\mu}$ (neglecting $\nu_e$), motivated by the fact that
only muon neutrinos are detectable as track events in neutrino
telescopes. Since oscillations in the neutrino sector mix the
different species, we chose instead to discuss the sum of all neutrino
flavors. When the effects of oscillations are accounted for, {\it
  nearly} equal numbers of the three neutrino flavors are expected at
Earth~\cite{Learned:1994wg}. (For a detailed calculation of the 
flavor ratios see e.g.,~\cite{Kashti:2005qa}.)

If the injected cosmic rays include nuclei heavier than protons, then
the neutrino flux expected from the cosmic ray sources may be
modified.  Nuclei undergoing acceleration can produce pions, just as
protons do, through interactions with the ambient gas, so the
Waxman-Bahcall argument would be unchanged in this case. However, if
interactions with radiation fields dominate over interactions with
matter, the neutrino flux would be suppressed if the cosmic rays are
heavy nuclei. This is because the photo-disintegration of nuclei
dominates over pion production at all but the very highest energies.
Defining $\kappa$ as the fraction of nuclei heavier than protons in
the observed cosmic ray spectrum, the resulting neutrino flux is then
given by
\begin{equation}
  E^2_{\nu} J_{\nu} \approx  (1 - \kappa) \,\,
[E^2_{\nu} \Phi_{\nu}]_{\rm{WB}} \,\,.
\label{simple}
\end{equation}
Unfortunately, current observations do not allow a conclusive
determination of $\kappa$. For $1.6 \leq \alpha \leq 2.1$, the data
can be well reproduced if the emitted cosmic rays consist entirely of
nuclei with masses in the intermediate (carbon, nitrogen, or oxygen)
to heavy (iron, silicon) range, but a mixture of protons (97\%) and
heavier species (3\%) is also
acceptable~\cite{Anchordoqui:2007tn,Anchordoqui:2007fi}.

\begin{figure}[tpb]
\rotatebox{0}{\resizebox{11.0cm}{!}{\includegraphics{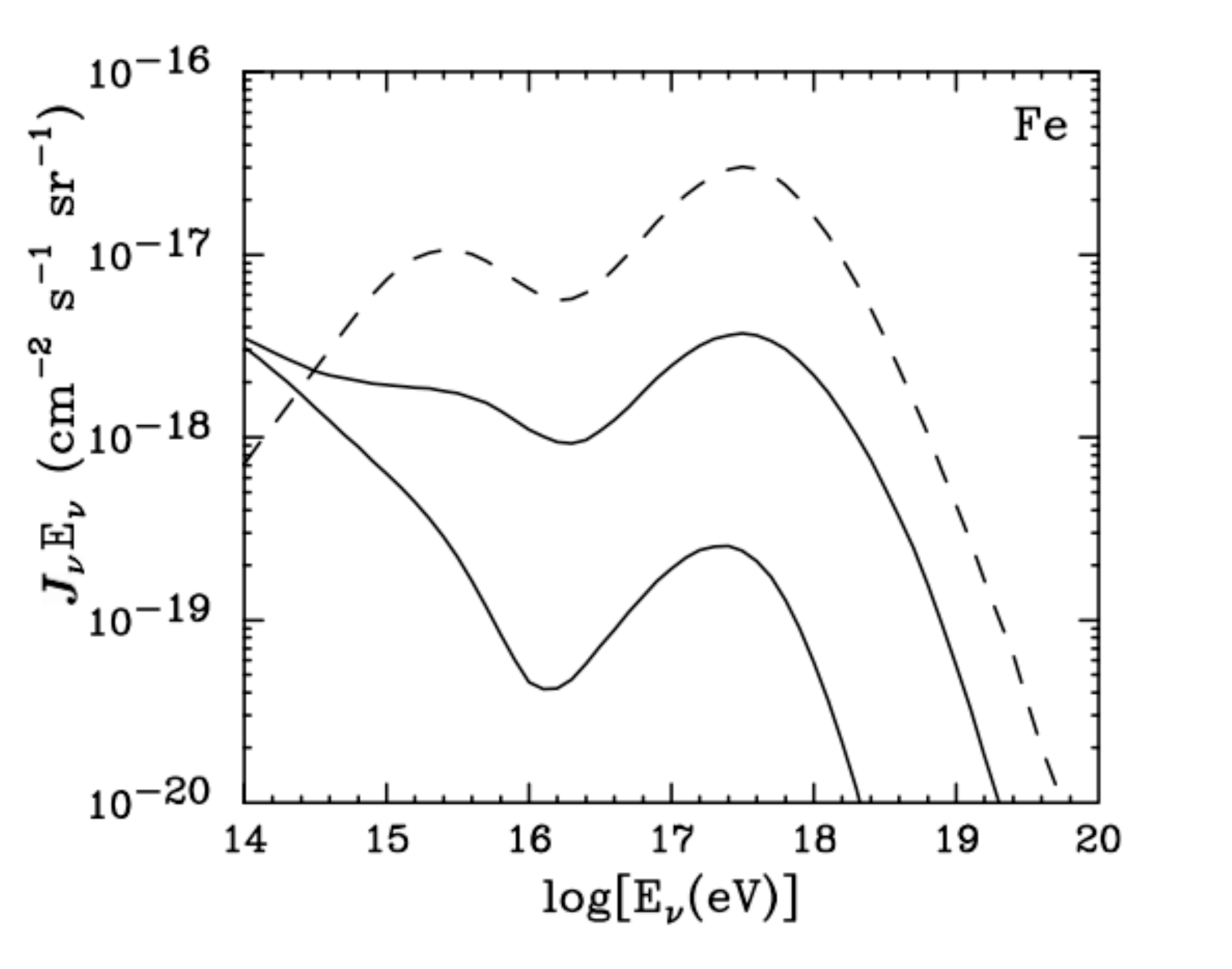}}}
\rotatebox{0}{\resizebox{10.5cm}{!}{\includegraphics{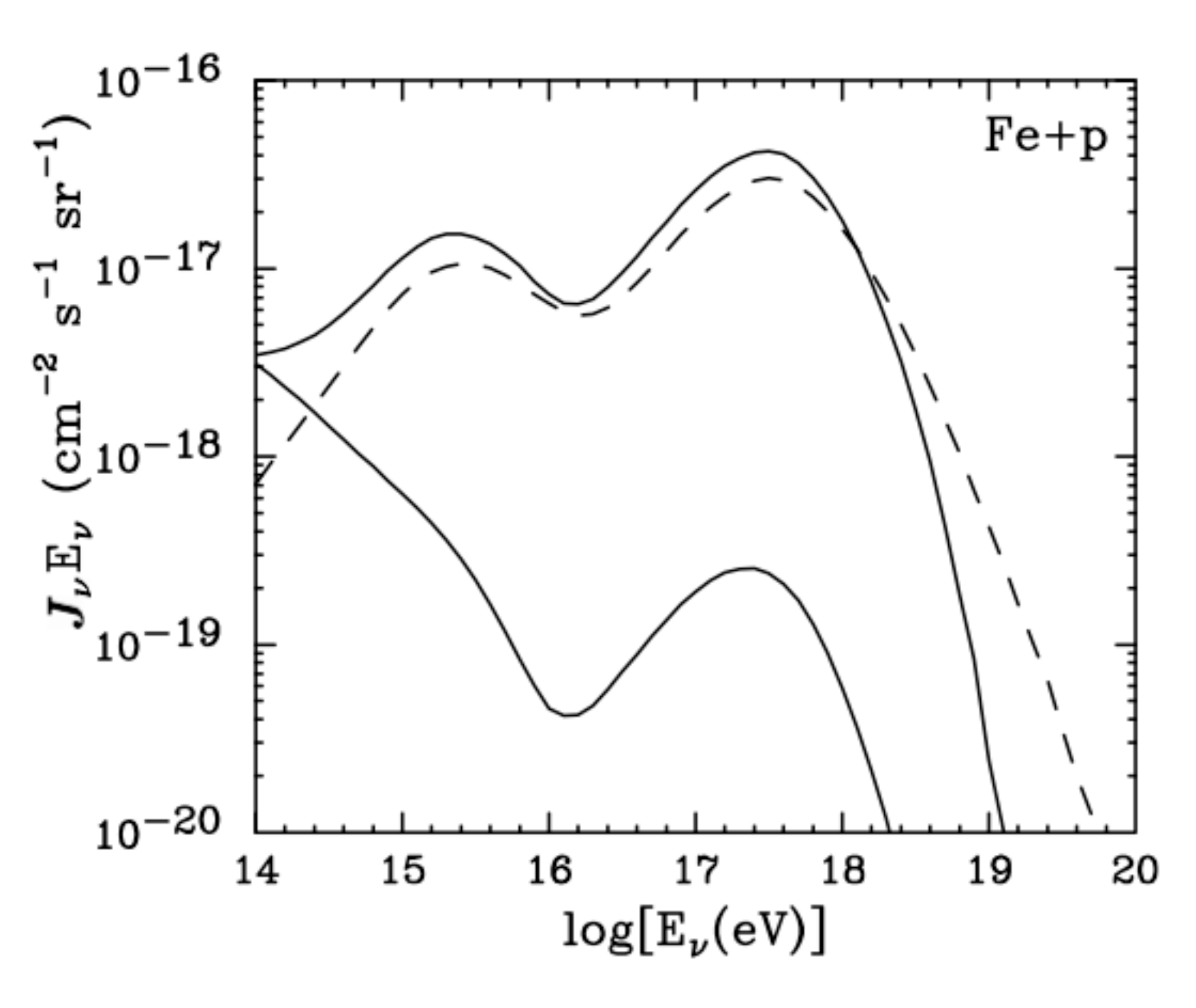}}}
\caption{The solid lines indicate the range for the cosmogenic
  neutrino spectrum for $\alpha=1.4-3.0$ and $E_{\max}/Z =
  10^{21}-10^{22}$~eV, for pure iron (up) and iron and proton (down).  
In each frame, we show for comparison as a
  dashed curve the prediction for an all-proton spectrum with
  $\alpha=2.2$ and $E_{\rm{max}} =10^{22}$~eV~\cite{Anchordoqui:2007fi}.}
\label{cosmogenic}
\end{figure}

The diffuse neutrino flux has an additional component originating in
the energy losses of ultra-high energy cosmic rays {\em en route} to
Earth. Ultrahigh energy protons above the ``GZK cutoff'' interact with
the cosmic microwave and infrared backgrounds as they propagate over
cosmological distances.  These interactions generate pions and
neutrons, which decay to produce neutrinos~\cite{cosmogenic}. The
accumulation of these neutrinos over cosmological time is known as the
cosmogenic neutrino flux.  Ultra-high energy nuclei also interact with
the cosmic microwave and infrared backgrounds, undergoing
photodisintegration. The disassociated nucleons then interact with the
cosmic microwave and infrared backgrounds to produce cosmogenic
neutrinos~\cite{Hooper:2004jc}. In the limit that the cosmic
backgrounds are opaque to cosmic ray nuclei, full disintegration
occurs and the resulting cosmogenic neutrino spectrum is not
dramatically different from that predicted in the all-proton case
(assuming the cosmic ray spectrum extends to high enough energies to
produce protons above the GZK cutoff). In contrast, if a significant
fraction of cosmic ray nuclei remain intact, the resulting flux of
cosmogenic neutrinos can be considerably suppressed. Of course, the
predicted neutrino flux also depends on the spectrum, the source
evolution, and the nature of the emitted cosmic
rays~\cite{cosmogenic,Engel:2001hd,Anchordoqui:2007fi}. Representative
spectra of cosmogenic neutrinos are shown in Fig.~\ref{cosmogenic},
for pure Fe and bi-modal Fe + p mass compositions consistent with
data~\cite{Anchordoqui:2007fi}.  The estimates shown in
Fig.~\ref{cosmogenic} are framed in the context of the usual
concordance cosmology~\cite{Amsler:2008zzb} of a flat universe
dominated by a cosmological constant with $\Omega_{\Lambda} \sim 0.7$,
the rest being cold dark matter (CDM) with $\Omega_\mathrm{m} \sim
0.3$. (A rigorous compilation of the $\Lambda$CDM parameters is provided
in Sec.~\ref{S_dark-matter}.) The Hubble parameter is given by
\mbox{$H^2 (z) = H^2_0\,(\Omega_{\mathrm{m} } (1 + z)^3 +
  \Omega_{\Lambda})$}, normalized to its value today of 70
km\,s$^{-1}$\,Mpc$^{-1}$. The time-dependence of the red-shift can be
expressed via $\mathrm{d}z = -\mathrm{d} t\,(1+z)H$. The cosmological
evolution of the source density per co-moving volume is parameterized
as
\begin{equation} 
\mathcal{L}_i(z,E) = \mathcal{H}(z)\mathcal{L}_i(0,E)\,,
\label{morton}
\end{equation}
where $\mathcal{H}(z)$ is the cosmological evolution of the cosmic ray
sources, which is taken here to follow the luminosity density
evolution of QSOs: $\mathcal{H}_{\rm QSO}  (z) = (1+z)^3$, for $z<
1.9$, $\mathcal{H}_{\rm QSO} (z) = (1 + 1.9)^3$ for $1.9 < z < 2.7$,
and $\mathcal{H}(z) = (1 + 1.9)^3 \exp\{(2.7-z)/2.7\}$, for $z >
2.7$~\cite{Engel:2001hd}. For such a cosmological evolution, which is
similar to that describing the star formation rate, $\xi_Z \approx
3$~\cite{Waxman:1998yy}.

In summary, the case for doing high energy neutrino astronomy is compelling; the challenge is to deliver the technology to build a neutrino detector with the largest possible effective area and the best possible angular and energy resolution sensitive to these cosmic neutrino fluxes. We discuss this next.

\section{NEUTRINO TELESCOPES}
\label{sec3}

\subsection{Detection principle}

\begin{figure}[tpb]
\rotatebox{0}{\resizebox{14.5cm}{!}{\includegraphics{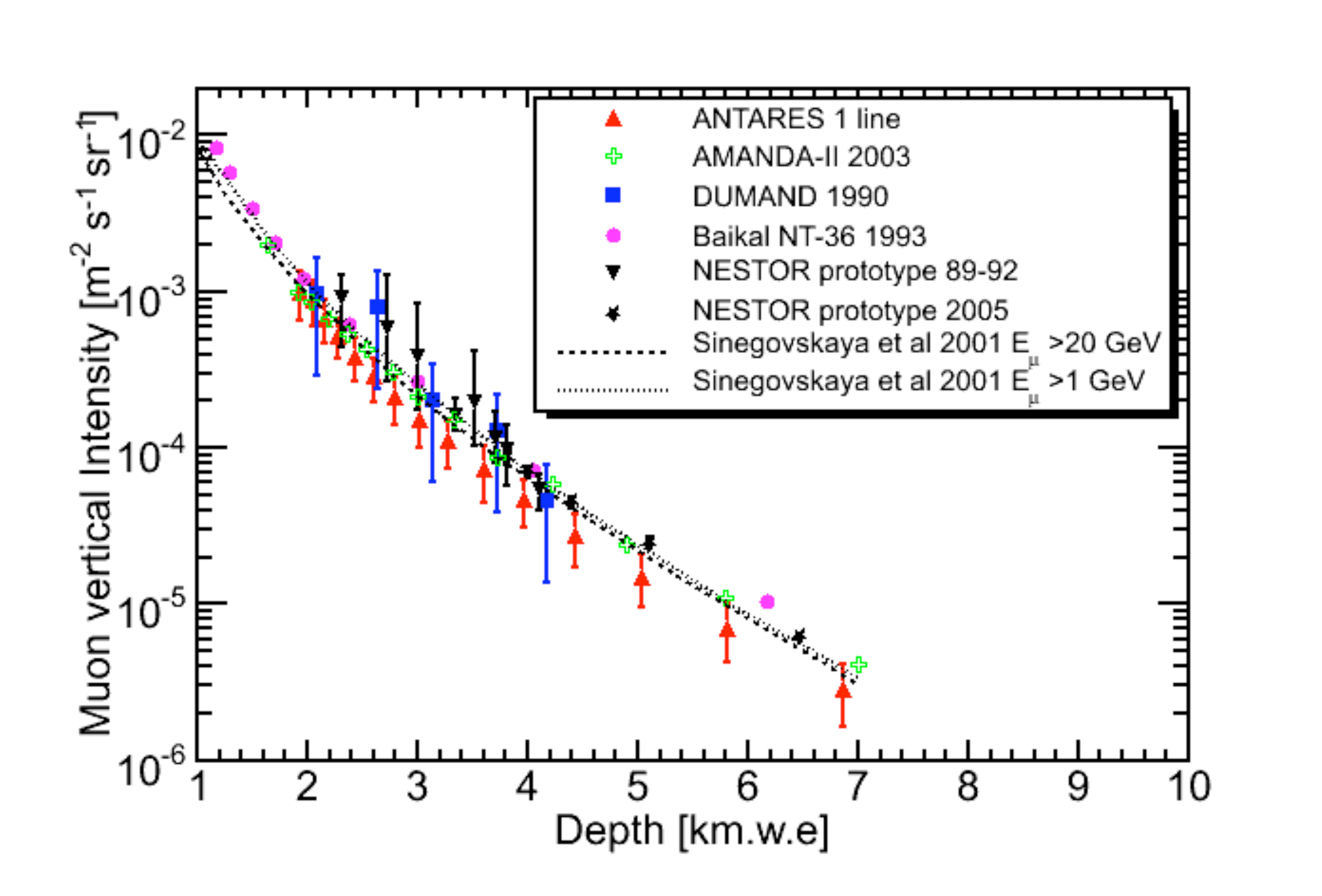}}}
\caption{Atmospheric muon vertical intensity measured by
  under-water and ice arrays as a function of depth compared to 
  calculations~\cite{Sinegovskaya:2000bv}. No correction is applied
  for the slightly different muon propagation properties  between
  water and ice.  Data are from: ANTARES and AMANDA-II~\cite{bazzotti}, Baikal,
  NT-36~\cite{Belolaptikov:1997ry}, the DUMAND prototype
  string~\cite{Babson:1989yy}, 
  and NESTOR (for 2 different prototypes)~\cite{Aggouras:2005bg}.}
\label{fig2a}
\end{figure}

\begin{figure}[tpb]
\rotatebox{0}{\resizebox{12.0cm}{!}{\includegraphics{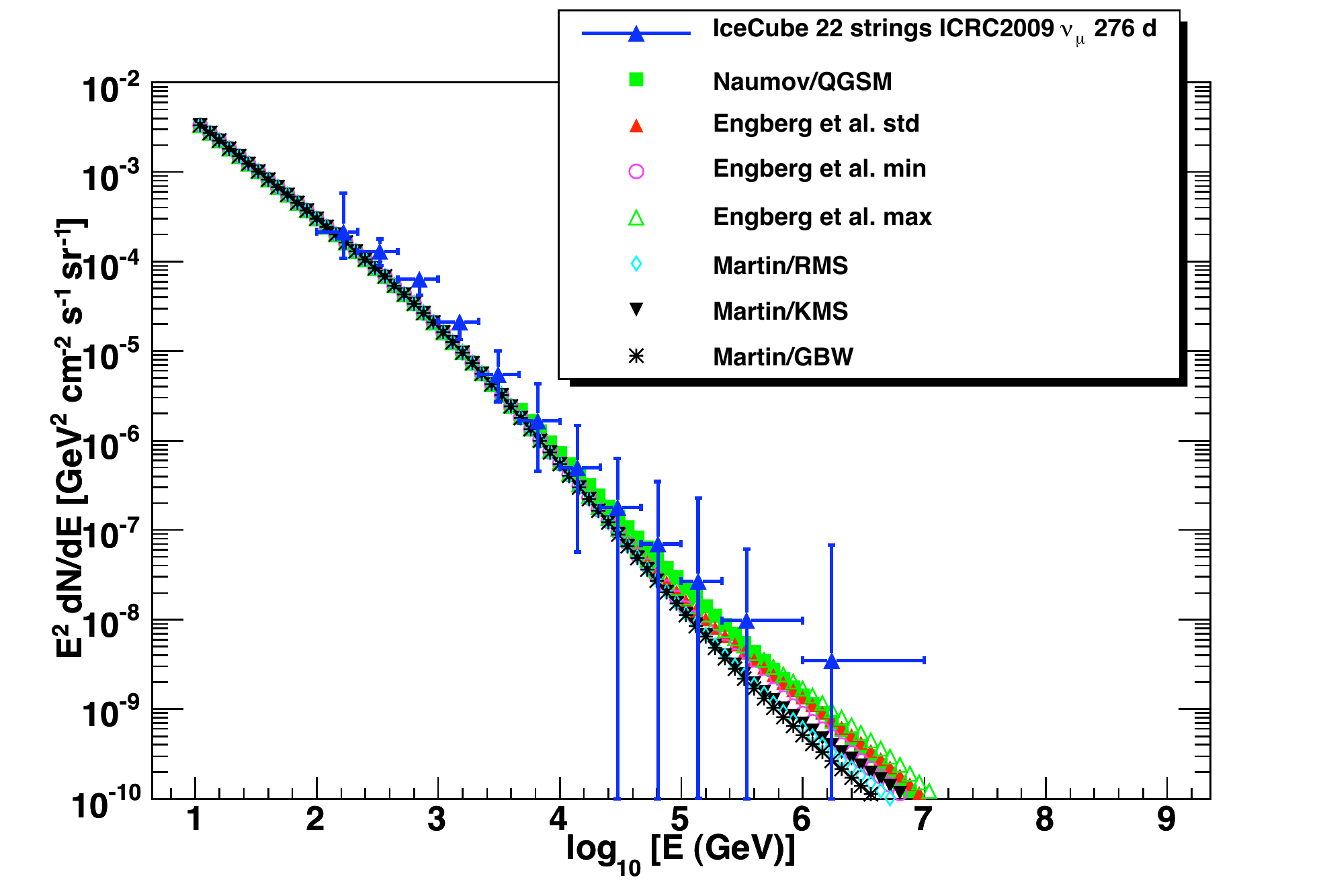}}}
\caption{Preliminary atmospheric unfolded neutrino spectrum measured
  with 22 strings of IceCube~\cite{dima} compared with predictions for
  $\nu_{\mu} + \bar{\nu}_{\mu}/3$. (The factor $1/3$ accounts for the
  lower CC interaction cross section of anti-neutrinos up to about
  $10^{5}$ GeV compared to neutrinos.) Data are accompanied by large
  error bars. A preliminary evaluation of  systematic errors, which are
 mainly due to the ice purity level dependence with depth, is
  included in the error bars.  The energy bin size was selected to
  match the energy resolution of the experiment.  (Since the energy
  resolution is 0.3 on the ${\rm log}_{10}(E_\mu),$ there are 3 bins
  per decade.) The size of the last 2 bins is larger, and determined
  by the desire to accommodate bins with vanishing contents.
  Conventional neutrino fluxes are calculated as
  in~\cite{Honda:2006qj}, prompt models are calculated in the
  framework of perturbative-QCD in~\cite{Enberg:2008te} (std stands
  for optimal parameters, whereas min and max indicate the range of
  variation of the model parameters) and in~\cite{Martin:2003us} for
  different structure functions; the
  Quark-Gluon String model (QGSM)~\cite{Fiorentini:2001wa} is also
  shown.}
\label{fig2b}
\end{figure}

It took about 35 years between Markov's conception of the Neutrino Telescope
(NT) detection principle~\cite{Markov:1960} to the operation of the
first full detector.  Though NTs use the well established
technique of photomultipliers (PMTs), they are challenging instruments
to build. The difficulties for construction arise from {\it (i)} the
necessity of instrumenting extremely large regions, mainly because of
the smallness of the neutrino-nucleon cross section and the steeply
decreasing energy spectrum of neutrino fluxes of atmospheric and
astrophysical origin; {\it (ii)} the required darkness and
transparency of the medium to detect the faint Cherenkov light
produced by ultra-relativistic charged particles; and {\it (iii)} the need
for filtering neutrino events out of the more abundant atmospheric
muons. Detectors are located under $1-4$~km of water or ice, where,
for $E_{\nu} \gtrsim 100$~GeV, the downward-going background of
atmospheric muons is about $5 - 6$ orders of magnitude larger than
the atmospheric neutrino flux coming from all directions (see
Figs.~\ref{fig2a}  and \ref{fig2b}).

NTs are tridimensional matrices of photodetectors made of strings of
optical modules (OMs) that house large photocathode PMTs and protect
them from both the water column pressure (when they are installed in
the sea or in lakes) and the pressure during ice refreezing (for
Antarctic detectors). In some cases OMs also comprise front-end
electronics. The NT energy threshold depends on the vertical spacing
between PMTs and the horizontal distance of strings; typical values
range between 50~GeV to 1~TeV. The threshold is not sharp, but the
detection efficiency increases with energy. Neutrinos interact with
the medium nuclei in and around the detector through charged
($\nu_{\rm lepton} + X \rightarrow {\rm lepton} + X$) and neutral
($\nu_{\rm lepton} + X \rightarrow \nu_{\rm lepton} + X$) current
interactions. Charged current deep inelastic collisions above 10~GeV
dominate over all other processes with the exception of $\bar \nu_e e$
interactions, because of the intermediate-boson resonance formed in
the neighborhood of $E_\nu^{\rm res} = M_W^2/2m_e \approx 6.3 \times
10^6~{\rm GeV}$, generally referred to as the Glashow
resonance~\cite{Glashow:1960zz}. The resonant reactions $\bar \nu_e
e^- \to W^- \to \bar \nu_\mu \mu^-$ and $\bar \nu_e e^- \to W^- \to$
hadrons may offer a detectable signal if the $\bar \nu_e$ flux
originates in $pp$ collisions~\cite{Anchordoqui:2004eb}.\footnote{In
  $pp$ collisions the nearly isotopically neutral mix of pions will
  create on decay a neutrino population in the ratio
  $N_{\nu_\mu}=N_{\overline\nu_\mu} = 2N_{\nu_e}=2N_{\overline\nu_e}.$
  In contrast, photopion interactions leave the isotopically
  asymmetric process $p\gamma\rightarrow \Delta^+\rightarrow \pi^+ n$
  as the dominant source of neutrinos; then, at production,
  $N_{\nu_\mu}=N_{\overline\nu_\mu} = N_{\nu_e} \gg
  N_{\overline\nu_e}.$}  The
neutrino-nucleon cross section rises almost linearly with
energy~\cite{Gandhi:1995tf}. This implies that {\it (i)} the
interaction probability increases with energy, indicating that NTs are
optimal detectors at high energies, and {\it (ii)} neutrinos begin to be
absorbed in the Earth. The shadowing effect of the Earth not only
depends on the neutrino energy but also on the incident zenith
angle~\cite{L'Abbate:2004hv}. At about $10^5$~GeV, the interaction
length of muon and electron neutrinos is about the Earth's diameter,
and hence about 80\% (40\%) of
the neutrinos with $\cos \theta = -1 \, (-0.7)$ are absorbed, where $\theta$ is
the zenith angle.  Hence, this indicates that at very high energy,
where it is expected that the astrophysical signal emerges from the
atmospheric muon and neutrino backgrounds, detectors should have a
large acceptance.  There is a completely different behavior for tau
neutrinos. In propagating through the Earth, they interact producing
tau-leptons that decay back into a $\nu_\tau$ of lower energy in what
is referred to as a ``regeneration effect''~\cite{Halzen:1998be}. The
two leptonic channels, with a branching fraction of about 17\%,
produce also $\nu_e$ and $\nu_\mu$.  The relevance of this phenomenon
in terms of event rates at detectors depends on the neutrino spectrum.
 
NTs detect ultra-relativistic ($\beta \sim 1$) secondary particles
produced in the interaction of the incoming neutrino with an atomic
nucleus. These particles, which typically travel faster than the speed
of light in the detector medium, emit directional Cherenkov light at
an angle $\theta_{C} = {\rm cos^{-1}}~[1/\beta n] \sim 41^{\circ}$,
where $n$ is the refraction index of the medium. Between 300-600~nm,
the region where PMTs are mostly sensitive, about $3.2 \times 10^4$
Cherenkov photons are emitted by a bare muon track in the ice per
meter. Once the OM acceptance and PMT quantum efficiency are accounted
for, on average only 6\% of these photons contribute to the
signal. The muon neutrino event topology is very different from that
of electron and tau neutrinos.  Secondary muons travel long distances
and can be reconstructed with an angular resolution better than a
degree at high energy.  Namely, the kinematic angle between the muon
and the neutrino in the deep inelastic regime ($\gtrsim 10$~GeV)
decreases with neutrino energy as $\sqrt{\langle \theta_{\nu \mu}^2
  \rangle} \propto \sqrt{m_{p}/E_{\nu}}$~\cite{Gaisser:1990vg}, with
$m_p$ the proton mass.  For $E_{\nu} \gtrsim 10$~TeV, the angle
becomes negligible with respect to the intrinsic detector resolution.
NTs measure the leading edge time of the PMT signal and in some cases
the fully digitized waveform and the integrated charge. This
information is used to reconstruct tracks for muons that travel long
distances across the detector and determine their energy as well.
Charged current (CC) interactions of electron and tau neutrinos and
all flavor neutral current (NC) interactions produce cascade-like
events. The resulting hadronic and electromagnetic showers are small
with respect to the scale of the separation between OMs, and so they
are almost a point-like source of light that propagates with some
anisotropy that remembers the direction of the incoming neutrino.
This prevents the achievement of an angular resolution for neutrinos
comparable to the muon channel.

\begin{figure}[tpb]
  \rotatebox{0}{\resizebox{13.5cm}{!}{\includegraphics{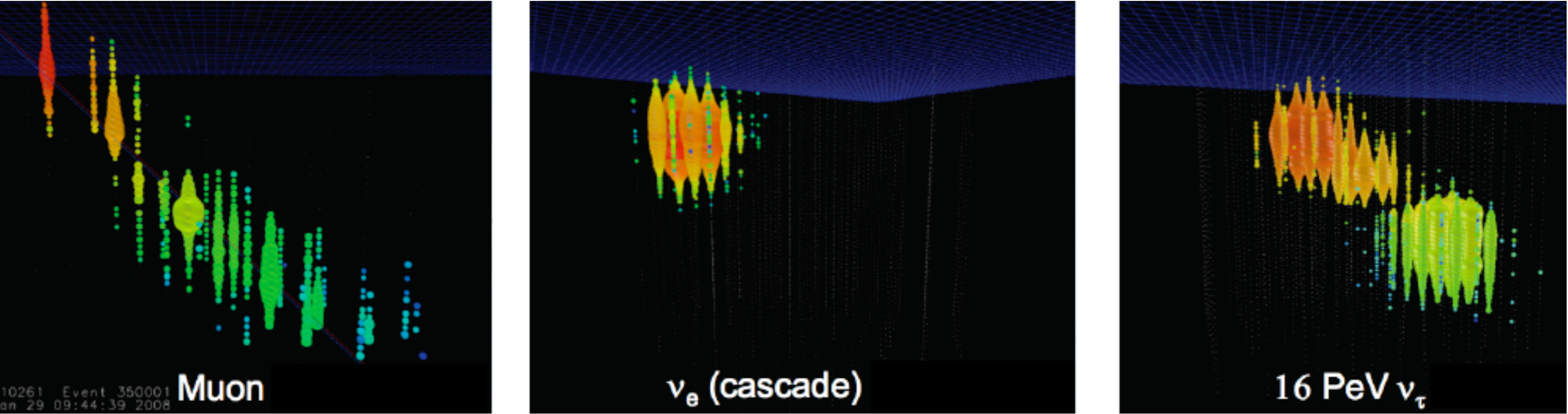}}} 
  \caption{Simulated events in the IceCube detector, visualized using the
    IceCube event display, showing the 3 typical topologies discussed
    in Sec 3. The shading represents the time sequence of the
    hits. The size of the dots corresponds to the number of photoelectrons
    detected by the individual photomultipliers. From left to right: a
    muon event of ~100~TeV, a cascade event induced by a ~100~TeV
    $\nu_e$, and a double bang event induced by a 16~PeV $\nu_\tau$.}
\label{fig:topologies}
\end{figure}

Muons lose energy according to 
\begin{equation}
\frac{dE_{\mu}}{dx} = - a - b E_{\mu},    
\end{equation}
where $a=2.0 \times 10^{-6}~{\rm TeV}\, {\rm cm}^2/{\rm g}$ is the continuous ionization term. The stochastic energy loss term $bE$, with
$b =4.2 \times 10^{-6}~{\rm cm}^2/{\rm g}$~\cite{Amsler:2008zzb}, dominates at energies $\gtrsim 1$~TeV in water or ice. The distance a muon travels before its energy
drops below some energy threshold $E^{\rm th}_\mu$, called the muon
range, is then given by
\begin{equation}  
\lambda_\mu = \frac{1}{b} \ln \left[ 
\frac{a + b E_\mu}{a + b E^{\rm th}_\mu} \right] \,.
\label{murange}
\end{equation}
For muons with $E_\mu > 1$~TeV propagating in water or ice, stochastic
losses, such as bremsstrahlung, pair production, and photonuclear
processes ($b E_{\mu}$ term) dominate over the ionization constant $a$
term.  In the first kilometer, a muon of about 100 TeV typically loses
energy in $2-3$ showers carrying more than 10\% of its initial
energy. Near the end of its range the muon becomes a minimum ionizing
particle emitting light that creates single photoelectron signals at a distance
of just over 10~m from the track. Above 1~TeV, the muon energy can be
reconstructed with an energy resolution in ${\rm log}_{10} E_{\mu}$ of
about 0.3. Of course, above a certain energy saturation effects of the
PMT electronics and the size of the detector limit the energy
reconstruction ability. In fact, because many of the events are not
fully contained at very high energy, part of the Cherenkov cone can
miss the PMTs so that part of the charge, that is proportional to the
muon energy, is lost. The parent neutrino energy can be inferred from
simulation of the $\nu (\bar{\nu}) + N \rightarrow \mu^{-} (\mu^{+})$
interactions. For $E_{\nu,\bar \nu} > 10^5$~GeV, the average value of
the energy taken by the lepton is $\sim
0.75$~\cite{Gandhi:1995tf}.  In most cases, the well reconstructed
cascade events develop inside the detector and hence the neutrino
energy resolution is largely improved, up to 0.1 - 0.2 in ${\rm
  log}_{10} E_{\mu}$. Tau neutrinos produce similar events to 
electron neutrinos up to about $10^6~{\rm GeV}$, because it is not yet
demonstrated that NTs have the ability to discriminate between
electromagnetic and hadronic showers. However, for $E_\nu \sim 2
\times 10^6~{\rm GeV}$, once again tau neutrinos deserve a special
attention: for such an energy, the tau-lepton range is about 100~m and
consequently both the $\nu_{\tau} N$ interaction and the shower
(hadronic or electromagnetic) from the tau decay can be
separated~\cite{Bugaev:2003sw}. These are called double-bang events
and are in principle a background-free topology. Nevertheless, the
expected event rates are limited by the small energy window in which
these events are detectable. As a matter of fact, this topology is
limited in energy to the range of a few PeV for the tau track to be
long enough, up to about 100~PeV to contain the two cascades in the
instrumented volume (the $\tau$ range is 1 km at about 200 PeV).  The
three different neutrino topologies are shown in
Fig.~\ref{fig:topologies}.

Atmospheric muons and neutrinos constitute a troublesome background for searches of neutrinos produced in astrophysical sources.  Moreover, the predictions for such a background are quite uncertain at energies $\gtrsim 1$~TeV. Therefore, background discrimination at NTs requires different techniques depending on the analysis. However, all of these analyses use the fact that neutrinos can cross the entire Earth, though with a shadowing effect that depends on energy and nadir angle.  By selecting upgoing muons induced by neutrinos, the residual background of atmospheric muons is largely reduced to those events that are downgoing but are mis-reconstructed as upgoing.  For cascade-events background discrimination is instead accomplished through containment cuts, since neutrinos have their vertex in the instrumented region while external tracks are produced mostly by atmospheric muons.

For point source searches, the directional information is the crucial
cipher. Being undeflected by intergalactic and galactic magnetic
fields, neutrinos preserve directionality from their own sources and tend to cluster around it while the
background distribution is random (see Sec.~\ref{ps}).  The golden channel to point back
to sources is the muon one. This is because the angular resolution is
better than $1^{\circ}$, whereas cascade like events have poor
resolution of about $10^{\circ}-30^{\circ}$, depending on the medium.

For diffuse flux searches, the energy measurement is determinant. This
is because the signal spectrum is expected to be harder than the
neutrino background~\cite{Gaisser:2002jj} from atmospheric cascades
initiated by cosmic rays. When protons and nuclei enter the
atmosphere, they collide with the air molecules and produce all kind
of secondary particles, which in turn interact or decay or propagate
to the ground, depending on their intrinsic properties and energies.
Above 1 GeV, the most abundant particles at sea level are neutrinos
and muons~\cite{Amsler:2008zzb}. The properties of these particles,
which constitute the main background for astrophysical searches, are
described in Sec.~\ref{atm-neu}.

\subsection{Detectors in operation and decommissioned}

The DUMAND and the Baikal collaborations pioneered underwater NT
technologies in the deep ocean near Hawaii~\cite{DUMAND:1980} and in
the Siberian Lake Baikal~\cite{Baikal:1984}, respectively. In November
1987, the DUMAND Collaboration measured the muon vertical intensity at
depths ranging between $2-4$~km, with a prototype line at a depth of
4.8~km about 30 km off-shore  of
Hawaii~\cite{Babson:1989yy}.  The project was canceled in 1995. In
1993 a first configuration of 36 PMTs on 3 strings (NT-36) was
installed in  Lake Baikal in Siberia at the shallow depth of 1.1~km
and 3.6 km off-shore~\cite{Belolaptikov:1997ry}. The experiment was
later upgraded to larger configurations (NT-72-96-144). In 700~days of
effective livetime, 320 million muon events were detected, and the
first atmospheric neutrino measured by a neutrino telescope was
selected in this sample and reported in 1996. The configuration
of the experiment (NT-200) with 192 OMs was put into operation in April 1998. It
consists of an umbrella-like structure of 8 (72~m long) strings, with
up-looking and down-looking OMs (containing 37-cm diameter
PMTs developed for the project, called QUASAR-370). Three
external strings at 100~m from the center of NT200, each with 12 pairs
of OMs, were added in April 2005 to increase the cascade
sensitivity at very high energies (NT200+). The Baikal Collaboration
is working towards the construction of a Gigaton-volume detector with
about 2100-2500 OMs (arranged over 95-100 strings separated by
distances of the order of 80-120~m) grouped in clusters to form
independent sub-arrays \cite{baikal_icrc09}. A prototype string has
been taking data since April 2008.  

The NESTOR Collaboration began
surveys of an area close to the Peloponnese coast at a mean depth of
about 4~km in the '90s.  They measured the muon flux with a hexagonal
prototype floor with 12 up-looking and down-looking
PMTs~\cite{Aggouras:2005bg}, and studied by simulation the possibility
to deploy a telescope made of towers with 12 of these hexagonal floors
(32~m in diameter) vertically spaced by $20-30$~m. The Collaboration is
operating a deep sea station off-shore of Pylos at a depth exceeding 3.5~km
connected to the on shore laboratory by an electro-optical cable.

In 1987 Halzen came out with the idea of ``ice-fishing''  neutrinos.                         
The Antarctic Muon And Neutrino Detector Array (AMANDA) was located
below the surface of the Antarctic ice sheet at the geographic South
Pole~\cite{Halzen:1992fi}.  During 1993 and 1994, in an exploratory
phase, the four-string $\rm AMANDA-A$ array was deployed and
instrumented with 80 PMTs spaced at 10~m intervals from 810 to
1000~m. (The scattering length at that depth is too short to allow
degree level reconstruction.) A deeper array of 10 strings, referred
to as AMANDA-B10, was deployed during the austral summers between 1995
and 1997, to depths between 1500 and 2000~m. The instrumented volume of
AMANDA-B10 formed a cylinder with diameter 120~m, overlooked by 302
optical modules (Hamamatsu R5912-2 20~cm diameter)~\cite{Andres:1999hm}. During December
1997 and January 2000, the detector was expanded by adding nine outer
strings of PMTs. The composite AMANDA-II array of 19 strings and 677
OMs forming two concentric cylinders with larger diameter of 200~m
became operative in 2000 and continued up to 2009. 
The OMs were connected to cables supplying the 8-inch PMT power and
transmitting analog signals. Some of the strings used fiber optic cables
reducing cross-talk effects, but most used coaxial twisted-pair
cables. One of the strings pioneered the digitization technique, with
data acquisition for waveform digitization in each module
communicating to the surface, designed as a prototype for the IceCube Digital
OMs (DOMs) and data transmission. 

\begin{figure}[tpb]
\rotatebox{0}{\resizebox{12.5cm}{!}{\includegraphics{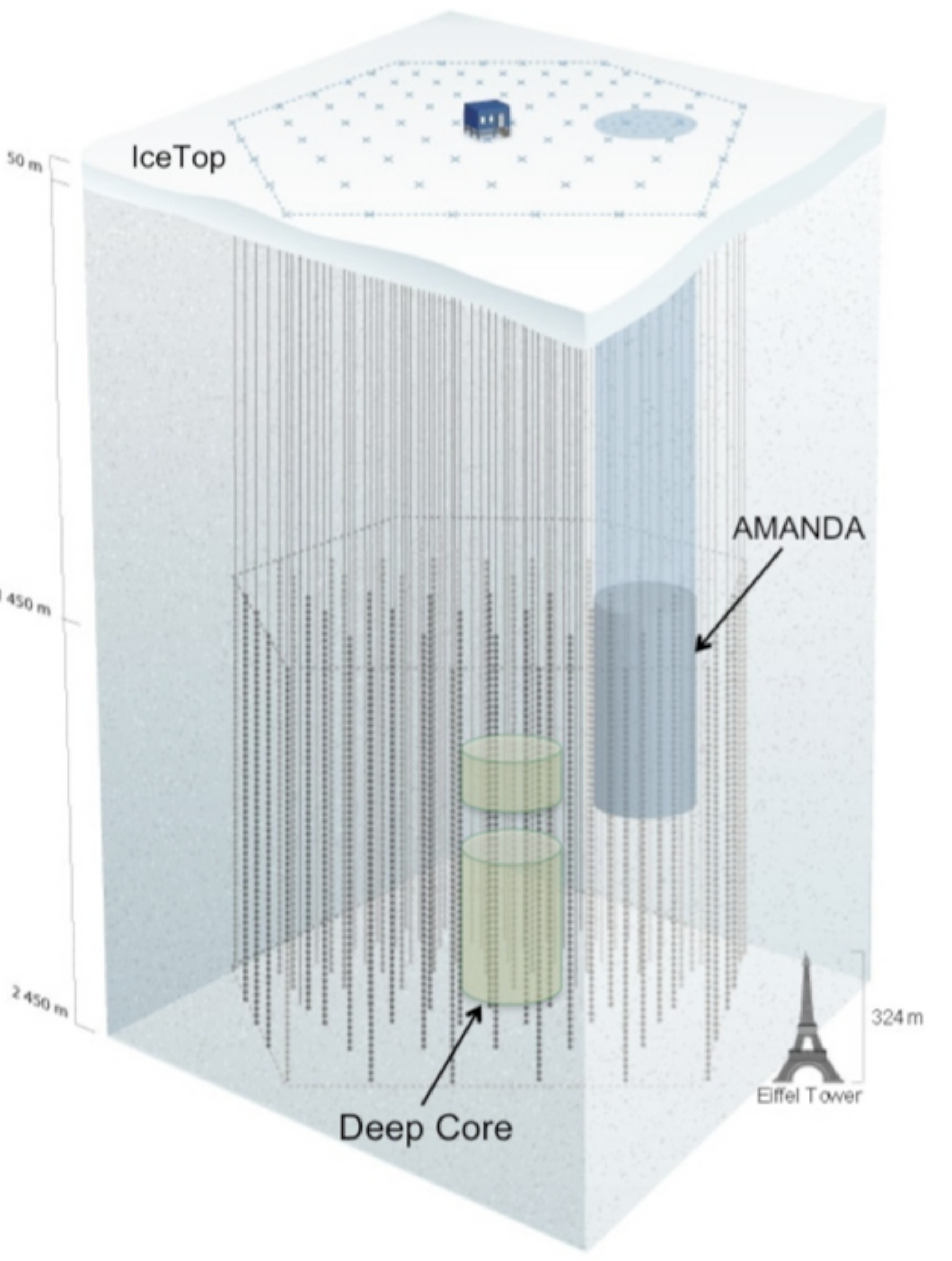}}}
\caption{The IceCube Observtory, including the deep ice array, IceTop,
  AMANDA, and Deep Core. For comparing sizes the image of the Eiffel
  tower is also shown.}
\label{icecube}
\end{figure}

The IceCube neutrino telescope, currently being deployed near the
Amundsen-Scott station, comprises a cubic-kilometer of ultra-clear ice
about a mile below the South Pole surface, instrumented with long
strings of sensitive photon detectors which record light produced when
neutrinos interact in the ice. The deep array is complemented by
IceTop, a surface air shower detector consisting of a set of 160
frozen water tanks. With its surface array, IceCube becomes a
3-dimensional air shower array for studies of cosmic rays up to
EeV. It is also useful for calibration of the neutrino telescope, and
it serves as a partial veto for the downward events at high energy.
IceCube strings are placed 124~m apart on a hexagonal grid. Each
string is instrumented (between 1.5 to 2.5~km below the surface) with
60 DOMs vertically separated by about 17~m. The DOMs enclose 10-inch
Hamamatsu PMTs with quantum efficiency $\sim 20\%$ at around 400~nm.
Six additional strings with new high efficiency ($\sim 25-30$\% at
400~nm) PMTs to be deployed in between 7 standard IceCube strings will
constitute the Deep Core.  It will improve the sensitivity of the
detector in the region $\lesssim 1$~TeV, important for dark matter
searches and neutrino oscillations.  The outer IceCube strings around
Deep Core and the upper layers of DOMs will be used as veto-shield
against the background of atmospheric muons allowing the
identification downgoing neutrino induced muons when the neutrino
vertex is in the inner detector. Construction of the observatory will
be completed in the austral summer 2010-11. The drilling of strings
has been so successful that up to about 20 strings can be installed in
a season (from mid-November to mid-February, when airplane flights are
possible to and from the South Pole).  The basic element of IceCube,
the DOM, is made of a glass spherical pressure vessel of 33~cm
diameter, containing a 25~cm diameter PMT, a high voltage supply,
light emitting diodes for energy and time calibrations, and an
electronic motherboard. The motherboard digitizes and time-stamps PMT
signals~\cite{:2008ym}.  The performance of the first string (deployed
in the 2005-6 season), its calibration and test methods, as well as
the data acquisition and installation of the detector are described
in~\cite{Achterberg:2006md}. The layout of IceCube is shown in
Fig.~\ref{icecube}.

\begin{figure}[tpb]
\begin{center}
\rotatebox{0}{\resizebox{10.cm}{!}{\includegraphics{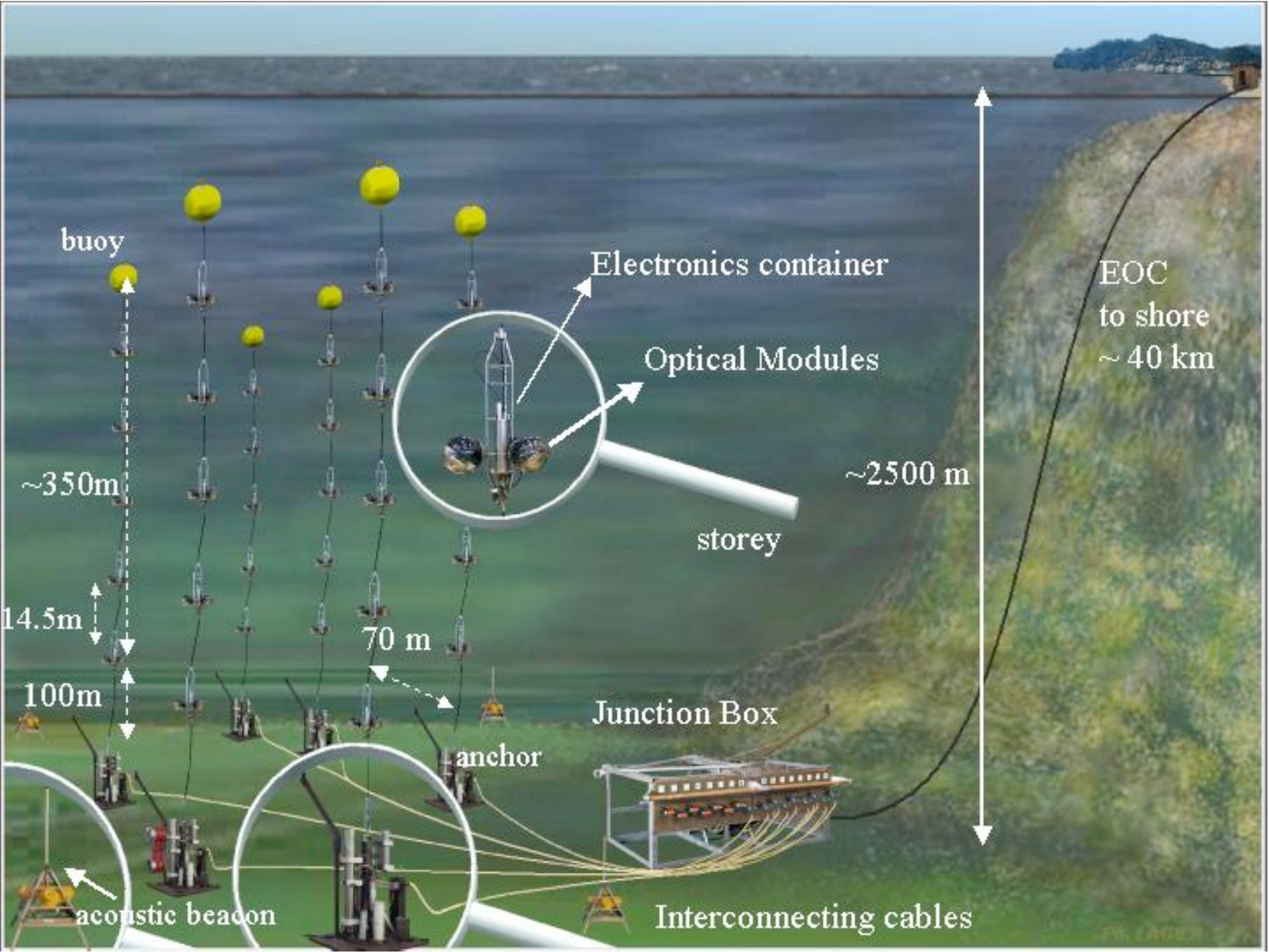}}}
\caption{A pictorial view of the ANTARES detector including an
 insert of a storey.}
\label{antares}
\end{center}
\end{figure}

The completion of the undersea neutrino observatory ANTARES, located
at about 2.5~km below the sea surface off-shore Toulon in South
France, took several years. Since 1996, the Collaboration had sea
campaigns comprising more than 60 line deployments for tests of water
transmission, light optical background, bio-deposition and
sedimentation~\cite{Amram:2002qt}. The permanent electro-optical
cable, of about 40~km, transmitting data and power between the shore
station in La Seyne sur Mer and the junction box off shore has been
operating since November 2002. Between 2003 and 2005 various prototype
lines and a mini-instrumentation line for environmental parameter
measurements were deployed~\cite{milom}. In 2006, the first 2 lines of
the detector were deployed and 8 additional lines were installed in
2007.  On May 2008 ANTARES was completed with a total of 12
lines. Lines were connected initially by a submarine and then by an
unmanned Remote Operated Vehicle. The ANTARES observatory comprises
the 12 mooring lines, a line specifically dedicated to marine
environmental monitoring, a seismometer, and a biocamera for
bioluminescence studies.  The lines are anchored to sea bed at 2475~m
depth and held vertical by buoys. Buoys are freely floating so each
line moves under the effect of the sea current, with movements of a
few meters for typical values of 5~cm/s. An acoustic positioning
system, made of transponders and receivers, gives a real time
measurement of the position of the OMs with a precision better than
10~cm, typically every 2 minutes. Tiltmeters and compasses provide
their orientation.  Seventy-five OMs along each lines between about
2400 and 2000~m are grouped in triplets on storeys. The 3 PMTs
(Hamamatsu 10"~\cite{pmt}) look downward at $45^{\circ}$ from the
vertical to prevent transparency loss due to sedimentation and
optimize detection for upgoing tracks. PMTs are housed inside pressure
resistant glass spheres made of two halves closed by applying an
under-pressure of 200-300~mbar. The set-up, including the PMT, the
glass sphere, the silicon gel for optical coupling between the glass
and the photocathode, and the mu-metal cage for shielding the Earth
magnetic field is referred to as the OM~\cite{om}. Storeys also
include titanium containers housing the front-end electronics, with a
pair of ASIC chips per PMT used for signal processing and
digitization. This provides the time stamp and the amplitude of the
PMT signal. Each of the OMs contains a pulsed LED for calibration of
the relative variations of PMT transit time and a system of LED and
laser Optical Beacons allows the relative time calibration of
different OMs.  An internal clock system, which is synchronized by GPS
to the Universal Time with a precision of $\sim 100$~ns, distributes
the 20 MHz clock signal from the shore. Time calibrations allow a
precision at the level of 0.5~ns ensuring the capability of achieving
an angular resolution at the level of $0.3^{\circ}$ muons above 10 TeV
for point source searches ~\cite{calibration}. All data above a
threshold of about 1/3 of a photoelectron pulse is sent to shore for
further online filter. This requires coincidences between PMTs on the
same storey rejecting hits not compatible to the propagation of light
in water between hit PMTs~\cite{daq}.
The ANTARES detector is shown in Fig.~\ref{antares}.

Ice and water are proving to be suitable transmission media with
different pros and cons. Photon propagation properties are
characterized by a much longer scattering length $\lambda_{\rm scatt}$
in sea water compared to ice. The effective scattering length,  defined
in terms of the average scattering angle of the photons, $\theta$, reads
$\lambda_{\rm eff} =  \lambda_{\rm scatt}/(1-\langle\cos\theta\rangle)$.
In sea water, $\lambda_{\rm eff} \sim 100-200$~m is much larger than in ice
due to a more forward peaked angular distribution. On the other
hand the absorption length, that determines the distance photons can
propagate before being absorbed, is much larger in ice ($\sim 100-200$~m)
than in water $\sim 50-70$~m.

Sea water is a less quiet environment than ice. Additionally,
bioluminescence and $\beta$-decay of $^{40}K$ can produce a baseline
of optical background in the sea of the order of $30-100$~kHz and MHz
bursts that cause a random dead time for OMs.  No current or
bioluminescence are present in ice. Once water developed in the
drilling process is frozen, the risk of short circuits due to water
leakages in the OMs is largely reduced, and PMTs can be switched
on. On the other hand, once ice freezes, the components are not
recoverable, while entire instrumented lines can be recovered from
water for repairs. The average background rate in IceCube PMTs is
$\sim 280$~Hz, mainly dependent on the radioactivity from the OM and
PMT materials.  The lower optical noise in ice makes IceCube a very
sensitive detector for SN collapse searches.\footnote{About 99\% of
  the binding energy of a Type II SN is expected to be released in
  neutrinos. These SNe originate in the gravitational collapse of
  massive giant stars, heavier than 8 solar masses, into a neutron
  star. Current upper limits on the number of supernova explosions in
  our Galaxy, from scintillator and water Cherenkov detectors, are
  getting closer to the predicted rates of about 2 per
  century~\cite{Montaruli:2009rr}.  In the early phase ($\sim 10$~ms),
  called deleptonization or neutronization, electron neutrinos are
  emitted in the electron capture process ($e^{-} + p \rightarrow n +
  \nu_{e}$) and subsequently ($\sim 10$~s) all neutrino flavors are
  produced in reactions such as $e^{+} + e^{-} \rightarrow \nu +
  \bar{\nu}$ in the thermalization phase. The passage of a large flux
  of MeV-energy neutrinos during a period of seconds will be detected
  as an excess of single counting rates in all individual optical
  modules. Each OM will be triggered by positrons from the
  inverse-$\beta$ reaction: $\bar{\nu}_{e} + p \rightarrow e^{+} + n$,
  where the visible positron energy is given by $E_{\rm vis} =
  E_{\bar{\nu}_{e}} - Q + m_{e} = E_{\bar{\nu}_{e}} -
  0.789$~MeV~\cite{Halzen:1995ex}. IceCube, with 5160 optical modules,
  can observe the 10~ms pulse as an increased rate over the average
  background rate (280~Hz) in the PMTs. For example, a SN at a
  distance of 7.5~kpc would produce a rate of about $10^6$~Hz over
  this time, yielding a 34$\sigma$ effect. A
  SN in the Large Magellanic Cloud would produce a $ 5\sigma$
  effect.}

\begin{figure}[tpb]
\rotatebox{0}{\resizebox{12.5cm}{!}{\includegraphics{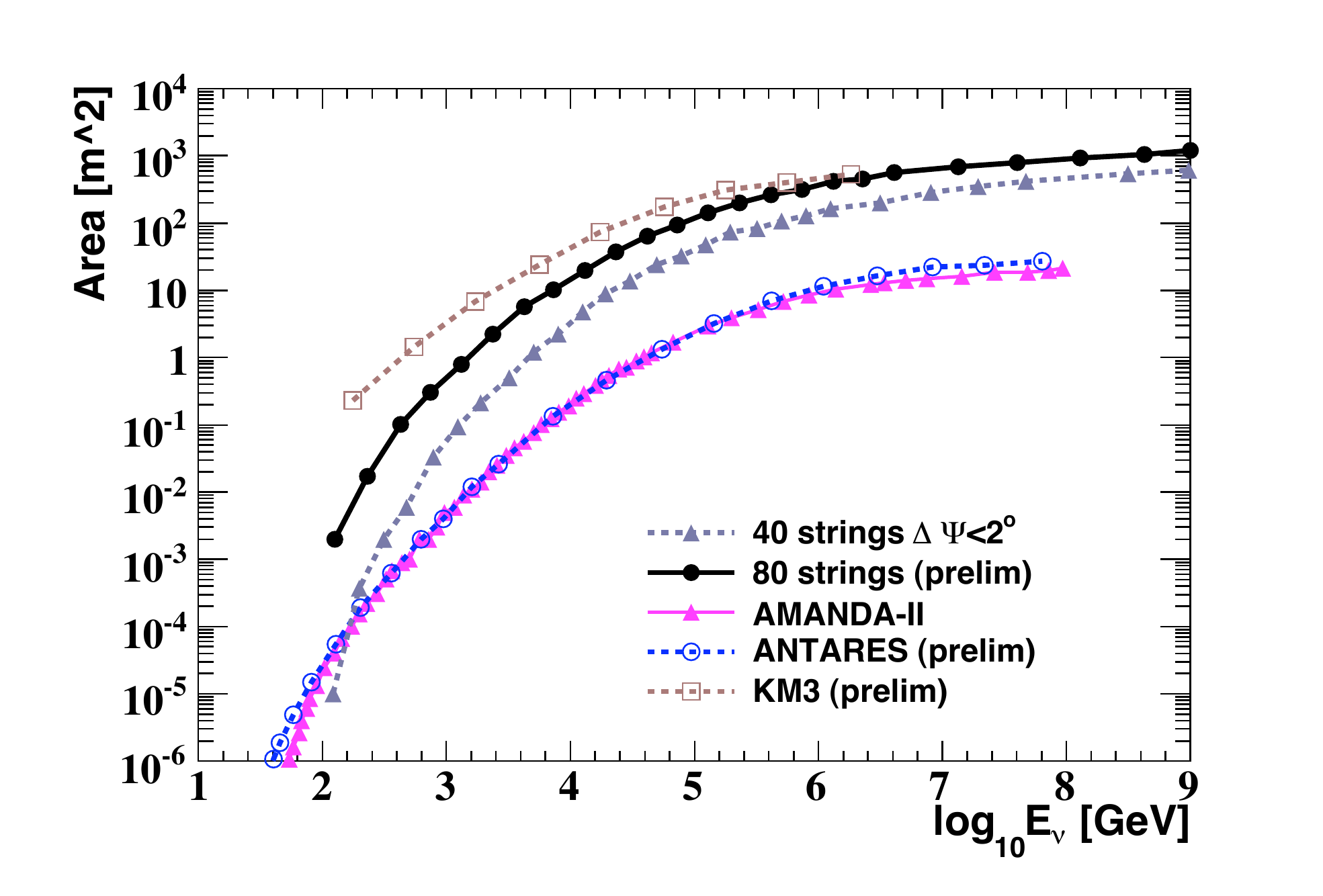}}}
\caption{Effective area for muon neutrinos, integrated over the
  lower hemisphere, as a function of the neutrino energy. The selected
  cuts ensure a pointing capability suitable for point source
  searches. The area for 40 strings of IceCube is obtained with the
  requirement that tracks are reconstructed inside $2^{\circ}$ from a
  source. The full IceCube area is obtained using the same cuts, and
  is therefore preliminary.  The AMANDA-II area is from the final
  analysis in Ref.~\cite{Abbasi:2008ih}. The ANTARES effective area is
  obtained with preliminary selection cuts based on
  simulation~\cite{Montaruli:2009kv}. The KM3NeT area is for a
  possible configuration still under debate~\cite{Sapienza:2009zz}.}
\label{performance1}
\end{figure}

The neutrino effective area, the sensitive area `seen' by
neutrinos producing detectable muons when entering the Earth, is a
useful parameter to determine event rates and the performance of a
detector for different analyses. The expected event rate 
\begin{equation}
N_{\mu} = \int\int \int
dE_{\nu} \ d\cos \theta_\nu \ d\phi_\nu \  A^{\rm eff}_{\nu}(E_{\nu}, \theta_\nu, \phi_\nu) \
J_\nu(E_{\nu}, \theta_\nu, \phi_\nu) \,,
\end{equation}
of a model predicting a neutrino spectrum $J_{\nu}$ depends on the neutrino-nucleon cross section $\sigma(E_\nu)$,
the shadowing of the Earth $P_{\rm Earth}(E_{\nu},\theta_{\nu})$, the
track reconstruction quality cuts, and the selection criteria for
background rejection. 
 The shadowing factor depends on column depth, that is
the integral of the Earth density profile $\rho (\theta_\nu, \ell)$ for a given
direction of the neutrino $\theta_\nu$ over the
distance in the Earth $\ell$:
$P_{\rm Earth}(E_{\nu},\theta_{\nu}) = e^{-N_A \times \sigma(E_{\nu}) \times
\int_{\ell} \rho(\theta_\nu,\ell) d\ell}$, where $N_A$ is the Avogadro number.
The neutrino effective area is a parameter calculable through simulations and
formally given by
\begin{equation}
A^{\rm eff}_{\nu}(E_\nu, \Omega_{\nu}) = V_{\rm eff}(E_\nu, \theta_\nu, \phi_\nu) \times
(\rho N_{A}) \times \sigma(E_{\nu}) \times P_{Earth}(E_\nu, \theta_{\nu})
\end{equation}
where $(\rho N_A)$ is the number of nucleons in the target material
assumed isoscalar and $V_{\rm eff}(E_\nu, \theta_\nu, phi_\nu) =
V_{\rm gen}(E_\nu, \theta_\nu, \phi_\nu) \times N_{\rm sel}(E_\nu,
\theta_\nu, \phi_\nu)/ N_{\rm gen}(E_\nu, \theta_\nu, \phi_\nu)$ is
the effective volume of the interactions generated in a volume $V_{\rm
  gen}$ of the neutrinos that produce detectable muons. The effective
volume depends on the energy of the neutrinos that determines how far
muons can travel from the vertex (muon range) and accounts for the
selection efficiency of an analysis (the efficiency factor $N_{\rm
  sel}/N_{\rm gen})$. The effective areas of the different NTs are
compared in Fig.~\ref{performance1}.

\begin{figure}[tpb]
\rotatebox{0}{\resizebox{12.5cm}{!}{\includegraphics{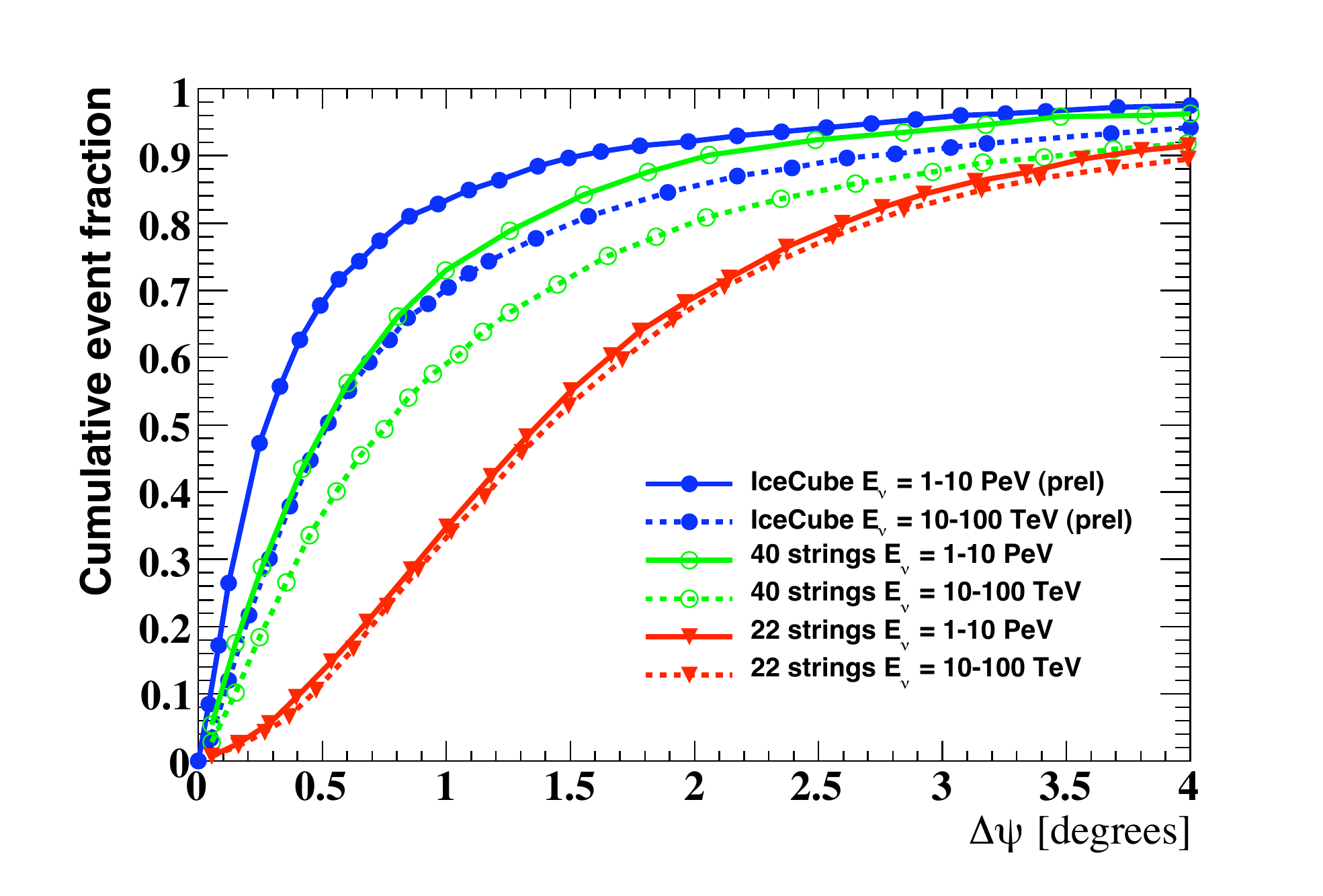}}}
\caption{Cumulative distribution of reconstructed muon events inside a
  cone of semi-angle $\Delta \Psi$ (Point Spread Function) from the
  neutrino direction. The different curves indicate growing
  configurations of IceCube for two neutrino energy bins and the
  point-source analysis cuts~\cite{Dumm}.  The IceCube curves are
  preliminary, because they are based on cuts optimized for the 40
  string analysis.}
\label{performance2}
\end{figure}

The point spread function (PSF) is the percentage of muons that are
reconstructed inside an angular half-cone width from the neutrino or
source direction. It depends strongly on the medium properties and on
the reconstruction algorithms.  In the final configuration of IceCube,
the angular resolution at energies between $10-100$~TeV will be such
that about 50\% of the reconstructed muons from the direction of a
neutrino point source will be inside an angle of $0.5^{\circ}$. Sea
water has longer effective scattering lengths than Antarctic ice hence
the resolution in ANTARES or in a future cubic-kilometer detector in
the Mediterranean~\cite{km3net} will be of the order of
$0.2^{\circ}-0.3^{\circ}$. The PSF of IceCube is shown in
Fig.~\ref{performance2}.

Cosmic ray experiments, like the Pierre Auger Observatory, provide a
complementary technique for ultra-high energy neutrino detection by
looking for deeply--developing, large zenith angle ($>60^\circ$) air
showers~\cite{Capelle:1998zz}. At these large angles, hadronic air
showers have traversed the equivalent of several times the depth of
the vertical atmosphere and their electromagnetic component has been
extinguished far away from the detector. Only very high energy
core--produced muons survive past 2 equivalent vertical
atmospheres. Therefore, the shape of a hadronic (background in this
case) shower front is very flat and very prompt in time. In contrast,
a neutrino shower appears pretty much as a ``normal'' shower. It is
therefore straightforward to distinguish neutrino induced events from
background hadronic showers. Moreover, very low $\nu_\tau$ fluxes
could be detected very efficiently by Auger detectors by looking at
the interaction in the Earth crust of quasi horizontal $\nu_\tau$
inducing a horizontal cascade at the detector~\cite{Bertou:2001vm}.

\section{EXPERIMENTAL RESULTS AND SCIENTIFIC IMPACT}
\label{sec4}

The main goal of neutrino telescopes is detecting astrophysical
neutrinos with $E_{\nu} \gtrsim 100$ GeV.  As we noticed in
Sec.~\ref{sec3}, muons constitute the ``golden sample'' for neutrino
astronomy since the achievable angular resolution is at sub-degree
level. At energies $\gtrsim 10$ TeV the limiting angle is mainly due to the
transit time spread of the PMTs and to the scattering of light.

While Imaging Atmospheric Cherenkov Telescopes (IACTs) can prove their
point spread function on copious sources of gammas, such as the Crab
Nebula, no astrophysical high energy neutrino source has proved to exist
yet. Hence, the `standard candle' for neutrino telescopes must be an
`anti-source', such as the Moon shadow. The Moon disk, with a diameter
of about $0.5^{\circ}$, blocks primary cosmic rays hence producing a
deficit in the muon flux. IceCube, using 3 months of data of the 40
string configuration, begins to find evidence at $5\sigma$ level of
this deficit, which is a proof of the absolute timing and pointing
capability of the detector~\cite{moon}. A detector with an ice
platform on top (like IceCube) is able to test, for a limited solid
angle, the angular resolution using events in coincidence with the
extensive air shower array IceTop. Sea detectors, such as ANTARES, can
use instead a scintillator array carried by a boat, or the
reconstruction of the shape of the close-by continental crust using
horizontal muons~\cite{Borriello:2009zz}.

\subsection{Steady and flaring point-source searches}
\label{ps}

Searches of point-like sources have various advantages when compared
to searches of diffuse fluxes, as in the former one can use the
direction, the energy, and eventually the time distributions to reject
the background of atmospheric neutrinos. As a matter of fact, the
signal is expected to distribute around the source according to the
detector PSF, and to have a harder energy spectrum than the
background. Flaring sources as well as GRBs are particularly
interesting, because if the neutrinos are emitted in coincidence with
the gamma flare it is possible to dramatically reduce the background
of neutrinos over a short time window.  Since the background of the
experiment is flat in right ascension (which is directly connected to
the hour of the day), many `equivalent experiments' with only
background can be reproduced by scrambling the right ascension of
events and keeping fixed other information, such as the declination
and the energy proxy fixed. This has the advantage that data can be
used for estimating the background rather than using simulation, so
that resulting significances of the signal and the background are not
affected by a possible imperfect understanding of the detector, or
similarly the detector medium. Moreover, for searches of flares or
GRBs, data outside the short time window can be used to reproduce
`equivalent experimental' samples where no signal is present. These
techniques are similar to the off-on source techniques applied in
photon astronomy, where the background outside the source pixel region
is used to evaluate the significance of the observed source.  In the
many neutrino astronomy point-source searches~\cite{Ambrosio:2000yx},
binned methods that would look for clusters of events on top of the
background from specific directions or in the entire sky were
applied. The search bin was optimized depending on the expected signal
shape (typically assumed to be $\propto E_\nu^{-2}$) but to avoid
missing signal events at the border of bins, grids or bins had to be
moved, hence, penalizing the search with many trial factors.  Many
experiments~\cite{Abbasi:2009iv} use likelihood
methods~\cite{Braun:2008bg} that can be up to 40\% more sensitive
depending on the amount of information used. These methods use
probability distribution functions that maximize the discrimination
between the signal and background. The signal is characterized with
respect to the background by the source directional feature and by the
harder spectrum expected from astrophysical sources compared to
atmospheric neutrinos. For the case of time dependent sources, such as
periodical binaries like micro-quasars or flares from blazars,
periodicity assumptions or light curves from gamma, X-ray, and optical
telescopes can be used.

\begin{figure}[tpb]
\rotatebox{0}{\resizebox{12.5cm}{!}{\includegraphics{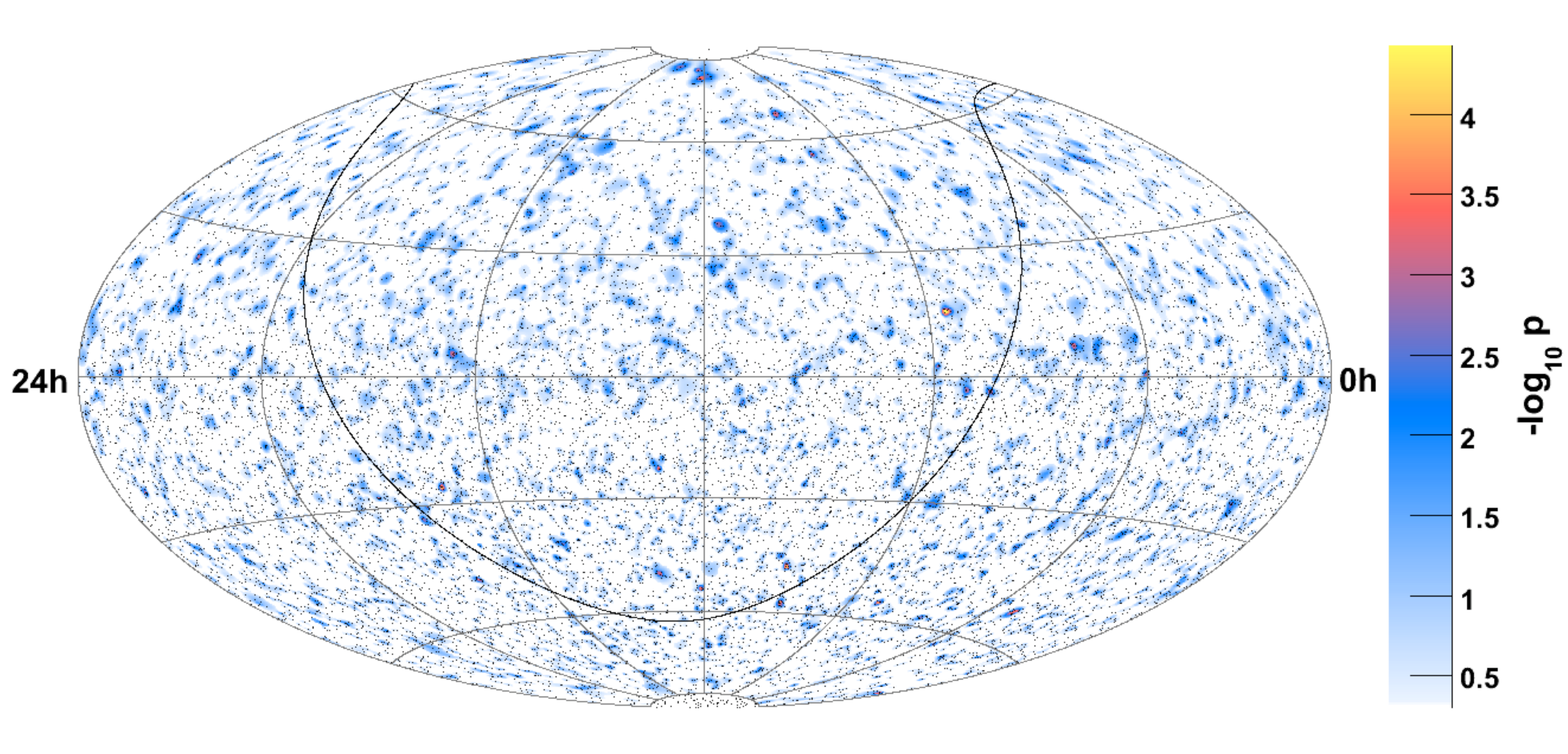}}}
\caption{Muon event skymap ($z$-axis color code is pre-trial
  significance) for six months of data taken with the 40-string
  IceCube array, from July 2008 through December 2008. Of the 17777
  black dots on the skymap, 6797 are upgoing neutrino candidates
  (zenith $> 90^\circ$) from the northern hemisphere. The color
  shading indicates the significance of the data and the curved black
  line is the galactic plane~\cite{Dumm}.}
\label{anisotropy1}
\end{figure}

The IceCube's 22 string configuration collected 5114 events from the
Northern hemisphere in 276 days~\cite{Abbasi:2008ih}.  No signal has
been found and upper limits have been set for sources in the Northern
hemisphere. For illustrative purposes we note that, for Geminga, the
Crab Nebula, Markarian 421, M87, and Cygnus OB2, the upper limits of
the 90\% CL interval for a flux $\Phi \leq \Phi_{90} \times
10^{-12}\mathrm{TeV}^{-1} \mathrm{cm}^{-2} \mathrm{s}^{-1} (E_\nu /
\mathrm{TeV})^{-2}$ are $\Phi_{90}= 9.67,\ 10.35,\ 14.35,\ 7.91,\
15.28,$ respectively. In addition, a hot spot was found with a
post-trial probability of 1.34\%.  The excess, however, was not
confirmed using the 6 month data of the 40 string configuration (that
by then encompassed in sensitivity the previous configuration by about
a factor of 2) and so the hot-spot has been considered a background
fluctuation~\cite{Dumm}. This 40 string analysis selects upgoing
neutrino events from the Northern hemisphere at a rate of about 40 per
day. In this region of the sky the experiment is sensitive mainly to
events in the $10-100$~TeV region for a $\propto E_\nu^{-2}$
spectrum. Very recently, by using scrambled background of very high
energy atmospheric muons, the IceCube Collaboration extended the field
of view for point source searches to the Southern hemisphere. In this
area of the sky, the sensitive region, for a spectrum $\propto
E_{\nu}^{-2}$, moves to higher energies
(PeV--EeV)~\cite{Abbasi:2009cv}.  As an illustration, the resulting
90\%CL upper limits $\Phi \leq \Phi_{90} \times 10^{-9}~{\rm GeV}^{-1}
\ {\rm cm}^{-2} \ {\rm s}^{-1} \ (E_\nu/{\rm GeV})^{-2}$, using 22
strings of IceCube, are  $\Phi_{90} = 558.2,\ 245.0,\ 27.1$, for
Centaurus A, Sagittarius A$^*$, and 3C279 respectively. (The
associated energy intervals that contain 90\% of the signal are:
$\Delta E_\nu/{\rm GeV} = 2 \times 10^6 - 8 \times 10^8,\ 1 \times
10^6 - 9 \times 10^8,\ 5 \times 10^4 - 1 \times 10^8$.) For the full
data sample of 40 strings, the IceCube discovery reach (5$\sigma$) for
$E_\nu^{2} \Phi_{90}$ (averaged over right ascension, and in the
declination region between $0^{\circ}-90^{\circ}$) is expected to be
between 1 to $5 \times 10^{-11}$~cm$^{-2}$ s$^{-1}$. This is near to
the exclusion flux for 22 strings and about a factor of 3 higher than
the exclusion limits for the same configuration.  With the 40 strings
result, flux limits for galactic sources are beginning to be in the
region of interest: $E_\nu^2 \Phi \sim 10^{-12}-10^{-11}$~TeV
cm$^{-2}$ s$^{-1}$ (see Sec.~\ref{sec2}). Actually, according to
estimates based on gamma-ray observations it would be possible to
achieve $5\sigma$ significance for Milagro Pevatrons, with $\sim
10$~yr of IceCube data collection~\cite{Anchordoqui:2006pb}. The 40
string sky-map is shown in Fig.~\ref{anisotropy1}.

\begin{figure}[tpb]
\rotatebox{0}{\resizebox{12.5cm}{!}{\includegraphics{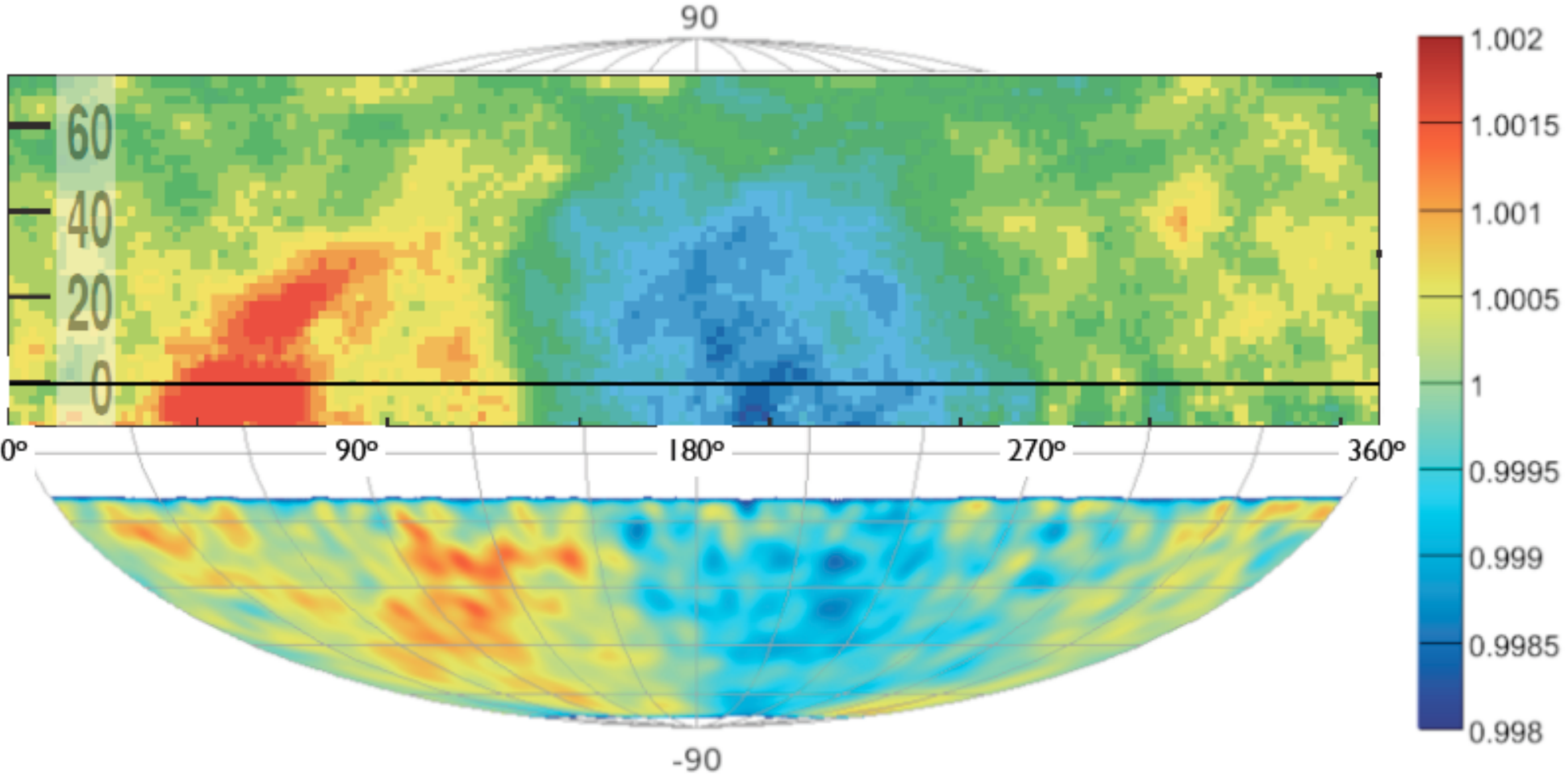}}}
\caption{Upper panel: Celestial
  cosmic ray intensity map for Tibet AS$\gamma$ data taken from
  2001-5. The vertical color bin width for the relative intensity, that is calculated 
  within each declination band,
  is $2.5 \times 10^{-4}$~\cite{Amenomori:2006bx}. The highest and lowest intensities are observed
  in the region of the tail-in and loss-cone in the direction of the heliotail, and on the right the excess is in the direction of Cygnus. Lower panel: Downward
  going muon skymap (expressed as relative intensity $\equiv$ number
  of events in an angular bin in right ascension and declination
  divided by the total number of events in the declination band)
  observed by 22 strings of the IceCube neutrino telescope with data
  taken between April 2006-7~\cite{Abbasi:2009wm}.}
\label{anisotropy2}
\end{figure}

\subsection{Atmospheric muons and neutrinos}
\label{atm-neu}

Atmospheric muons and neutrinos detected by NTs above some tens of
GeVs are produced in cosmic ray induced showers and come mainly from
pions and kaons. At lower energies ($\lesssim 10$~GeV) muon and
electron neutrinos are also generated by muon decays. They were
extensively studied in the past two decades because of the so called
``atmospheric neutrino problem,'' a deficit of muon induced neutrino
events that was explained by muon neutrino oscillations into tau
neutrinos [for a review see,
e.g.,~\cite{Gaisser:2002jj,GonzalezGarcia:2007ib}]. At energies above
100~GeV, where NTs operate, atmospheric neutrino oscillation studies
are not a primary target.  Nonetheless, although DeepCore and ANTARES
photocathode coverage are modest, they will collect a very large
statistics of atmospheric neutrinos opening the possibility of
observing neutrino oscillations. The main interests for NTs to measure
atmospheric neutrinos are {\it (i)} calibration of the detector,
demonstrating their ability to discriminate neutrino events from the
huge background of atmospheric muons; {\it (ii)} understanding of
hadronic interactions in the forward region, at energies not yet
accessible by colliders; and {\it (iii)} study of new physics scenarios.

The dominant production channels of atmospheric and muon neutrinos
detected in NTs are the decay chains of mesons created in hadronic
cascades. The pion decay chain ($\pi^+ \to \mu^+ \nu_\mu \to e^+ \nu_e
\nu_\mu \overline \nu_\mu$ and the conjugate process) dominates the
atmospheric neutrino production at $\sim$~GeV energies.  At energies of some GeVs, the muon decay length begins to
be comparable to the atmospheric altitude where most 
muons and neutrinos are produced ($10-20$~km). Hence, depending on the
incoming zenith angle, muons may stop decaying. At energies of about
100~GeV kaon decay becomes increasingly important with respect to pion
decay. In this energy range, the atmospheric neutrino spectrum can be
parametrized as~\cite{Gaisser:1990vg}
\begin{equation}
\frac{dN_{\nu_{\mu}}}{d {\rm ln}E_\nu} = A_{\rm tot}E_\nu^{\alpha} \left[ \frac{A_{\nu}}{1+B_{p}E_\nu (\cos\theta)^{*}/\epsilon_{\pi}} + 0.635 \times \frac{B_{\nu}}{1+B_{k}E_\nu (\cos \theta)^{*}/\epsilon_{k}} \right]
\label{eq:numu}
\end{equation}
where 0.635 is the branching ratio of $K^{\pm}$ decays;
$\epsilon_{\pi} = 115$~GeV and $\epsilon_{K} = 850$~GeV are the
critical energies respectively for pion and kaon decay, $\alpha \sim
1.7$ is the spectral index of the primary spectrum, $A_{\rm tot}$;
$A_{\nu}$ and $B_{\nu}$ are normalization constants; and $B_{p}$ and
$B_{k}$ are constant terms with energy.  The term $(\cos\theta)^{*}$
is equal to $\cos\theta$ for zenith angles such that $\cos\theta
\lesssim 0.3$, and for horizontal directions ($\cos\theta \gtrsim
0.3$) it has a larger value that accounts for the curvature of the
Earth. The $1/(E_\nu\cos\theta)$ dependence, is characteristic of
atmospheric processes where there is competition between interaction
and decay, since the decay probability increases with respect to the
interaction one for longer horizontal paths in the atmosphere. The
muon flux in this energy region has a similar dependence with the
important difference on the energy fraction taken by muons in two body
pion and kaon decays. We remind the reader that, on average, in pion
decay the neutrino takes a quarter of the energy of the pion while the
muon takes the rest.  This implies that above $\sim 1$~TeV, muon
production is dominated by pion decay and neutrino production by kaon
decay (about 85\% of the muons are produced via pion decay, whereas
only 25\% of neutrinos come from the decay of charged pions).  The
electron neutrino formula is more complicated, with two similar terms
due to $K^{\pm}$ and $K_{L}$ decays from which $\nu_{e}$ are produced
directly, and a term due to muon decay into electron neutrinos
determined by $\pi^{\pm}$, $K^{\pm}$ and $K_{L}$ decays. In this
energy window, the flavor ratios are $w_e : w_\mu : w_\tau \approx
1/20 : 19/20 : 0$~\cite{Lipari:1993hd} and the energy spectra are a
function of the zenith angle of the atmospheric cascades. As already
observed, mesons in inclined showers spend more time in rarefied
atmosphere where they are more likely to decay rather than interact.
For this reason the spectra of highly inclined neutrinos are harder
than those of almost vertical neutrinos and resemble their primary
spectrum. Above about $10^5$~GeV, kaons are also significantly
attenuated before decaying and the ``prompt'' component, arising
mainly from very short-live charmed mesons ($D^\pm,$ $D^0,$ $D_s$ and
$\Lambda_c$) dominates the spectrum~\cite{Zas:1992ci}. Such a
``prompt'' neutrino flux is isotropic with flavor ratios $12/25 :
12/25 : 1/25.$ Since prompt neutrinos have a hard spectrum resembling
the primary one (below the knee the differential cosmic ray spectrum
goes as $\sim E^{-2.7}$), they are a persisting background for
astrophysical neutrinos that follow a power law $\sim E_\nu^{-2}$.

Various Monte Carlo high precision calculations have been developed
for studying the atmospheric neutrino
oscillations~\cite{Barr:2004br,Honda:2006qj,Gaisser:2002jj}. For
energies below 10~GeV, high precision needs to account for the angle
of secondaries with respect to primaries~\cite{Battistoni:1999at}, the
solar modulation effects, and the Earth geomagnetic field
deflections. At higher energies these effects become negligible and
one dimensional calculations that neglect the angle of secondaries
respect to primaries are suitable. Uncertainties in atmospheric
neutrino fluxes are of the order of 15\% in the region of $\sim
10$~GeV, rising to about 20-30\% at $\sim
100$~GeV$-1$~TeV~\cite{Barr:2006it}. These are mainly due to the
rather poor knowledge of the primary cosmic ray flux and the hadronic
interaction models.  A precise estimate in the prompt region is not
possible since charm production in the atmosphere is not well
constrained by collider measurements~\cite{Berghaus:2007hp}. The
various calculations differ by up to 2 orders of
magnitude~\cite{Zas:1992ci}. Some of them are shown in
Fig.~\ref{fig2b}~\cite{Enberg:2008te,Fiorentini:2001wa,Martin:2003us}. The
experiments LHCf and TOTEM, dedicated to measure particles emitted in
the very forward region~\cite{D'Alessandro:2009zz}, will improve our
understanding of hadronic models at laboratory energies of about
$10^{17}$~eV.

A cubic-km NT measures of the order of kHz muon events, the so-called
``second calibration beam.''  With such a high statistics, high
precision verifications become possible; e.g., how well the detector
details and the effects of the light propagation in the radiator are
simulated.  Nonetheless, hadronic interaction models used in the
atmospheric shower simulations are affected by theoretical
uncertainties, not only experimental ones.  Some of these models are
similar to those used to simulate atmospheric neutrinos. This
represents a drawback, because the two calibration beams should be
used to understand the `same' detector independently of theoretical
uncertainties.  Another complication that affects shower models is the
difficulty of investigating separately errors coming from imperfect
knowledge of high energy primary cosmic ray composition and hadronic
model, and from the fact that simulations do not account for the
seasonal variation of the atmosphere. These simulations are very time
consuming since rates of events are very high and high energy showers
develop many particles in the atmosphere. Hadronic interaction models
that foresee a larger $K^+$ production (such as SIBYLL 2.1) show
better agreement than others (e.g. QGSJET II) in the high $x = E_{\rm
  secondary}/E_{\rm primary}$ region~\cite{Ahn:2009wx}.  The
preference for such hadronic models is also indicated by the muon
ratio measurement of MINOS~\cite{Adamson:2007ww} that shows a rise
between $0.3-1$~TeV consistent with an increasing contribution to the
muon charge ratio from kaons.

A neutrino telescope's high statistics measurement of the atmospheric
neutrino flux above about 100~GeV is an important benchmark for all
these calculations. A fit of the atmospheric muon neutrino flux [see
Eq.~(\ref{eq:numu})] was recently performed with AMANDA-II data to
determine variations with respect to existing
models~\cite{Collaboration:2009nf}. The allowed region in the
parameter space normalization vs spectral index was derived relative
to the Bartol flux~\cite{Barr:2004br}, using the angular distribution
and an energy proxy for 5511 neutrino events collected in 1387 d. The
best fit point indicates that, at 640~GeV, the data prefer a higher
normalization by about 10\% and a slightly harder spectrum by about
0.056. Using a sample of 4492 upgoing neutrino induced muon events
collected in 275.5 days, with a contamination of 5\% from
misreconstructed atmospheric muons, the IceCube Collaboration unfolded
the atmospheric neutrino spectrum for the 22 string
configuration~\cite{dima}. This spans a wider energy range up to PeV
energies and is compared to some conventional (from $\pi$ and $K$ decays)
and prompt neutrino models in Fig.~\ref{fig2b}.
Systematic errors are so large at $E_\nu> 10$~TeV that it is not yet
possible to disentangle an eventual contribution from prompt neutrinos
(originating in the decay of charmed mesons and baryons) from an
astrophysical component~\cite{Enberg:2008te}.

Neutrino telescopes can collect a huge statistics of muons, of the
order of billions.  Thus, they are sensitive to anisotropies of
amplitudes $< 10^{-3}$. For example, in Fig.~\ref{anisotropy2} we show
the map in equatorial coordinates of about 5 billion atmospheric muons
detected with the 22 strings of IceCube. These muons are produced by
cosmic rays with a median energy of $14$~TeV and have a median angular
resolution of $3^{\circ}$, and show an anisotropy in their arrival
direction of amplitude $(6.4 \pm 0.2) \times 10^{-4}$.  The observed
anisotropy persists with high statistical significance, to energy in
excess of 100 TeV with a smaller but measurable amplitude.  This
observation is not compatible with the Compton-Getting
effect~\cite{compton}, which produces an energy-independent dipole
anisotropy in the arrival direction of cosmic rays due to the relative
motion of the solar system with respect to the rest frame of galactic
cosmic rays (GCRs). This explanation was already excluded by Tibet
Array~\cite{Amenomori:2006bx} that in the absence of an anisotropy at
300~TeV concluded that GCRs co-rotate with the Galaxy. Other
experiments observed anisotropies in the Northern hemisphere at the
scale of $10^{\circ}-30^{\circ}$~\cite{Abdo:2008aw}. Particularly
Milagro applied a cut that discriminates hadrons with respect to
gammas.  The observed anisotropy with IceCube in the opposite
hemisphere persists at energies $> 100$~TeV, and hence makes it
difficult to explain it as an effect of the heliosphere or the
heliospheric tail.  Interestingly, it has been speculated that the
effect may be due to the propagation of cosmic rays produced by nearby
($\sim 100$~pc) SNRs and pulsars \cite{Salvati:2008dx}. It is
noteworthy the presence of Geminga between the anisotropy regions in
the Northern hemisphere. In the Southern hemisphere, the Vela pulsar
is located near the highest intensity regions observed by IceCube. Of
course such regions are so broad that we can only speculate about
responsible local sources in the absence of a more precise
calculation, including the intensity of galactic magnetic fields. This
in turn could be correlated to observations at lower energies of an
excess in the positrons to electron fraction with respect to secondary
production models~\cite{Adriani:2008zr} that have also been
interpreted as propagation of cosmic rays from local sources such as
SNRs and pulsars~\cite{Hooper:2008kg}. It has also been speculated
that the sites of the acceleration of cosmic rays may not be isolated,
but also involve superbubbles (overlapping SNRs)~\cite{Butt:2009zz}.
 
\subsection{Searches for  diffuse cosmic neutrino fluxes}

\begin{figure}[tpb]
\rotatebox{0}{\resizebox{12.9cm}{!}{\includegraphics{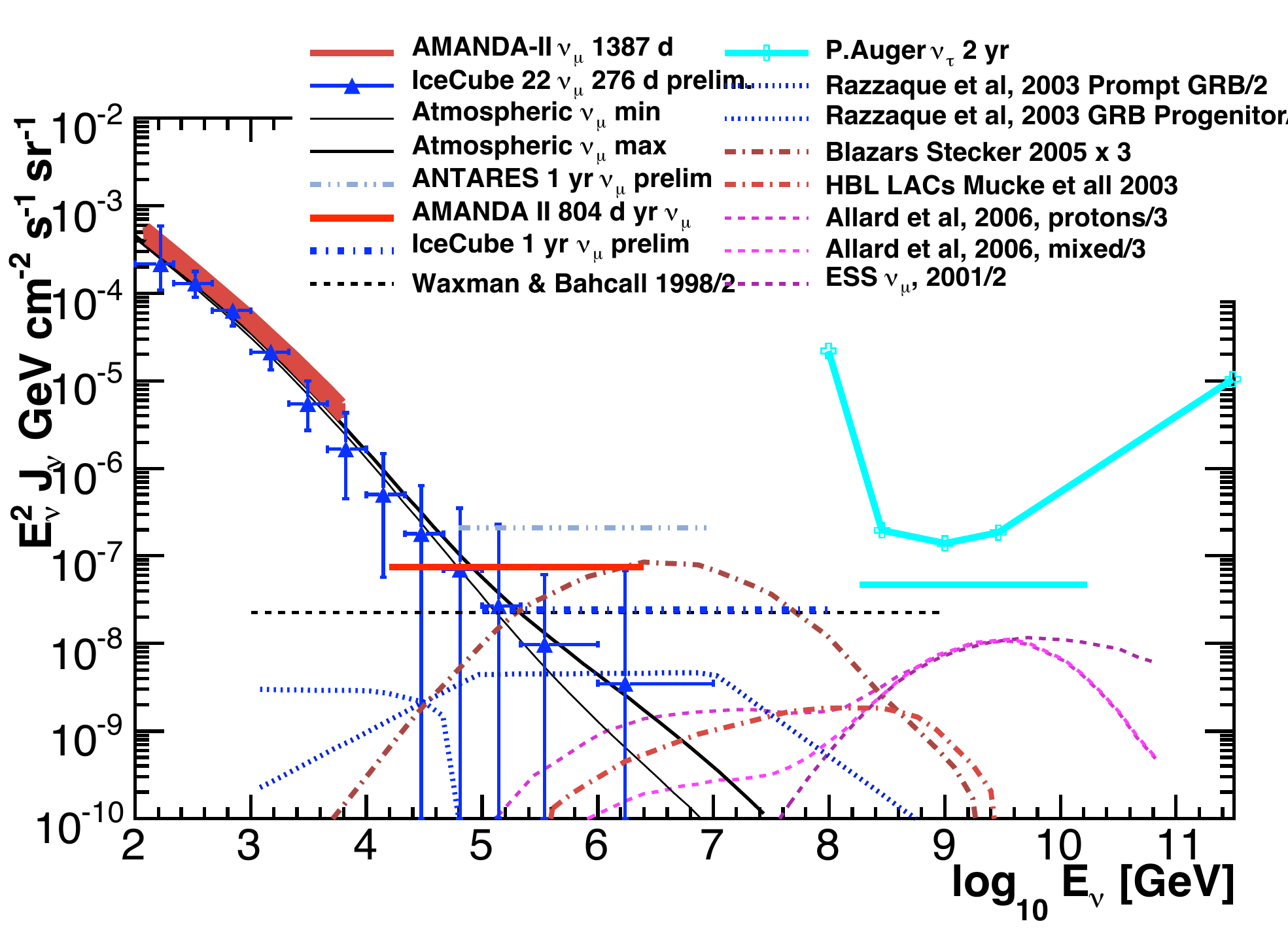}}}
\caption{The blue triangles indicate the atmospheric
  ($\nu_{\mu}+\bar{\nu}_{\mu}$) flux averaged over the lower
  hemisphere as measured by 22 strings of IceCube~\cite{dima}), and
  the falling thick solid line indicates AMANDA-II
  measurements~\cite{Collaboration:2009nf}).  For comparison,
  predictions of the atmospheric neutrino flux (conventional + prompt)
  are indicated by the falling thin solid
  lines~\cite{Barr:2004br,Honda:2006qj,Fiorentini:2001wa,Enberg:2008te}.
  The horizontal lines represent different 90\%CL limits and
  sensitivities on $E_\nu^{-2}$ fluxes; from top to bottom: ANTARES
  estimated sensitivity for 1~yr \protect\cite{coyle}, AMANDA-II muon
  neutrino limit for 804 days~\cite{Achterberg:2007qp}, IceCube
  estimated sensitivity for 1 yr~\cite{Ahrens:2003ix}. For comparison,
  also shown are {\it (i)} the Waxman and Bahcall upper limit
  corrected for oscillations~\cite{Waxman:1998yy}, {\it (ii)} neutrino
  flux predictions from AGNs~\cite{Stecker:1991vm} and prompt and
  precursor emission from GRBs~\cite{Waxman:1997ti}, and {\it (iii)}
  various extimates of the nearly guaranteed cosmogenic
  flux~\cite{cosmogenic,Engel:2001hd}. The ultra-high energy upper
  limit on the tau neutrino flux (differential and $\propto
  E_\nu^{-2}$) reported by the Pierre Auger Collaboration is also
  shown~\cite{Abraham:2009eh}.} 
\label{diffuse1}
\end{figure}

\begin{figure}[tpb]
\rotatebox{0}{\resizebox{12.5cm}{!}{\includegraphics{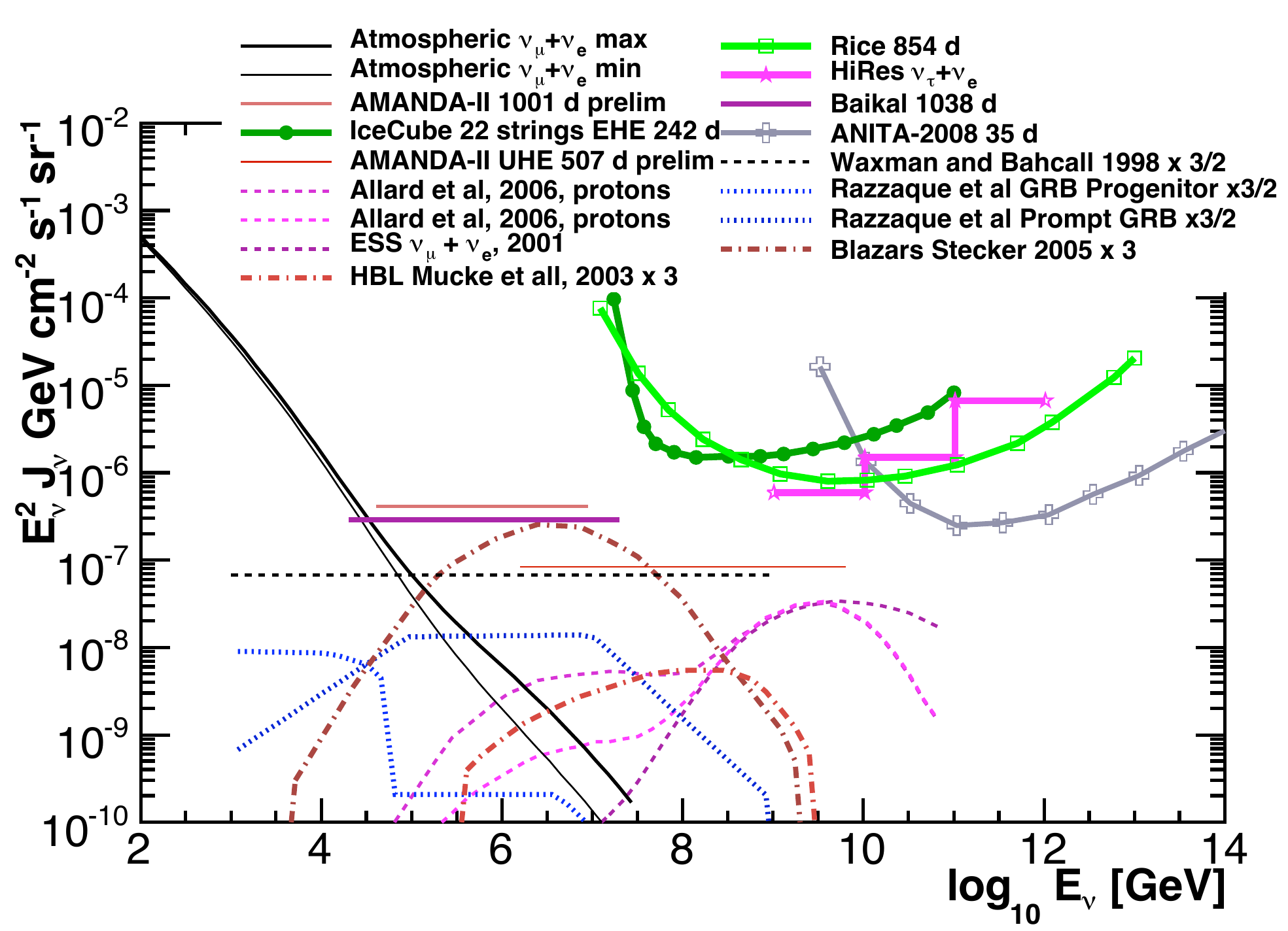}}}
\caption{ Same as in Fig.~\ref{diffuse1} but for experiments measuring
  more than one flavor, e.g. measuring cascades induced by all flavor
  neutrinos.  The horizontal solid lines are 90\%CL limits for a flux
  $\propto E_\nu^{-2}$; from top to bottom: AMANDA-II all-flavor
  cascade limit for 1001 days~\cite{amanda_cascades}, Baikal cascade
  limit for 1038 days~\cite{baikal_icrc09}, AMANDA-II muon neutrino
  limit for 804 days~\cite{Achterberg:2007qp}, all flavor ultra-high
  energy limit (preliminary) for 507 days of
  AMANDA-II~\cite{Silvestri}. The differential in energy upper limits
  (90\% CL) are for the 22 strings of IceCube~\cite{ic22_ehe} and
  several ultra-high energy neutrino detection experiments, including
  those with radio detectors that look for the coherent Cherenkov
  radiation produced by the excess of electrons in neutrino induced
  showers in dense media (e.g., ice and the lunar regolith). For
  further details see~\cite{Montaruli:2009rr}.}
\label{diffuse2}
\end{figure}

Diffuse fluxes from extra-galactic sources are more promising compared
to single source fluxes because of the larger event rates, with a
signal that may extend up to the highest energies where some of the
extra-galactic sources (e.g., as AGNs and GRBs) can
contribute. Nonetheless, these searches have to rely more on
simulations than point source searches and hence are subject to larger
systematic errors.  The basic concept is that because of the harder
spectrum the astrophysical signal should show up at high energies
above the atmospheric neutrino background. However, the theoretical
uncertainties on high energy atmospheric neutrino fluxes (see
Sec.~\ref{atm-neu}) and the experimental errors in the high energy
region where the statistics is low, limit the sensitivity and
reliability of this search. For large statistics, the uncertainty in
the prompt neutrino flux can be mitigated by triggering on tau
neutrinos, which have the double bang characteristic signature in the
PeV energy range, see Fig.~\ref{fig:topologies}. Specifically, the
detection of a double bang event would confirm neutrino oscillations
and at the same time establish the discovery of high energy cosmic
neutrinos; this is because tau neutrino production in the atmosphere
is negligible.  Figures~\ref{diffuse1} (experiments measuring 1
flavor) and \ref{diffuse2} (experiments measuring all neutrino
flavors) present a collection of experimental limits compared to
representative models for classes of sources, such as AGNs, GRBs, and
cosmogenic neutrinos. These upper limits can in turn be used to
constrain the origin and nature of ultra-high energy cosmic rays.  It
is this that we now turn to study.

\subsection{Cosmic ray proton fraction}

Consider the working hypothesis introduced in Sec.~\ref{sec2} for
optically thin sources, i.e., the relative number and energy of the
neutrinos and neutrons depend only on kinematics, which implies
approximate equipartition of the decaying pion's energy between the
neutrinos and the electron, and the relative radiation density in the
source. On average, each interaction will produce $\eta$ neutrinos per
neutron with relative energy $\epsilon$ per neutrino, {\it i.e.}  $
\eta = \langle N_\nu\rangle/\langle N_n\rangle$ and $\epsilon =
\langle E_\nu\rangle/\langle E_n\rangle.$ The neutrino emissivity of
flavor $i$ is then given by
\begin{equation}
\frac{\Delta E_{\nu_i}}{N_{\nu_i}}\mathcal{L}_{\nu_i}(z,E_{\nu_i})
= \frac{\Delta E_n}{N_n}\mathcal{L}_n(z,E_n)\,.
\end{equation}
Assuming flavor universality as well as $\epsilon \simeq E_\nu/E_n
\simeq \Delta E_\nu/\Delta E_n$ and $\eta \simeq
N_{\mathrm{all}\,\nu}/N_n$ one arrives at the neutrino source luminosity
(per co-moving volume)
\begin{equation}\label{ratio}
\mathcal{L}_{\mathrm{all}\,\nu}(z,E_\nu) \simeq\frac{\eta}{\epsilon}\,
\mathcal{L}_n(z,E_\nu/\epsilon)\,.
\end{equation}
For the (hypothetical) source where pion production proceeds
exclusively via resonant $p\gamma \to \Delta^+$ scattering, the
neutrino multiplicity and the relative energy are fixed: $\eta=3$ and
$\epsilon=0.07.$ Note, that the relation~(\ref{ratio}) derived for
optically thin sources can be regarded as a {\em lower} limit on the
neutrino luminosity as long as energy-loss processes in the source are
negligible~\cite{Ahlers:2009rf}.  Therefore, this conservative
expectation can be translated into an {\em upper} limit on the
extra-galactic proton fraction in ultra-high energy cosmic rays,
exploiting, e.g., the experimental upper limit on the diffuse neutrino
flux from IceCube or the preliminary AMANDA data~\cite{Silvestri}. 

\begin{figure}[tpb]
\rotatebox{0}{\resizebox{10.5cm}{!}{\includegraphics{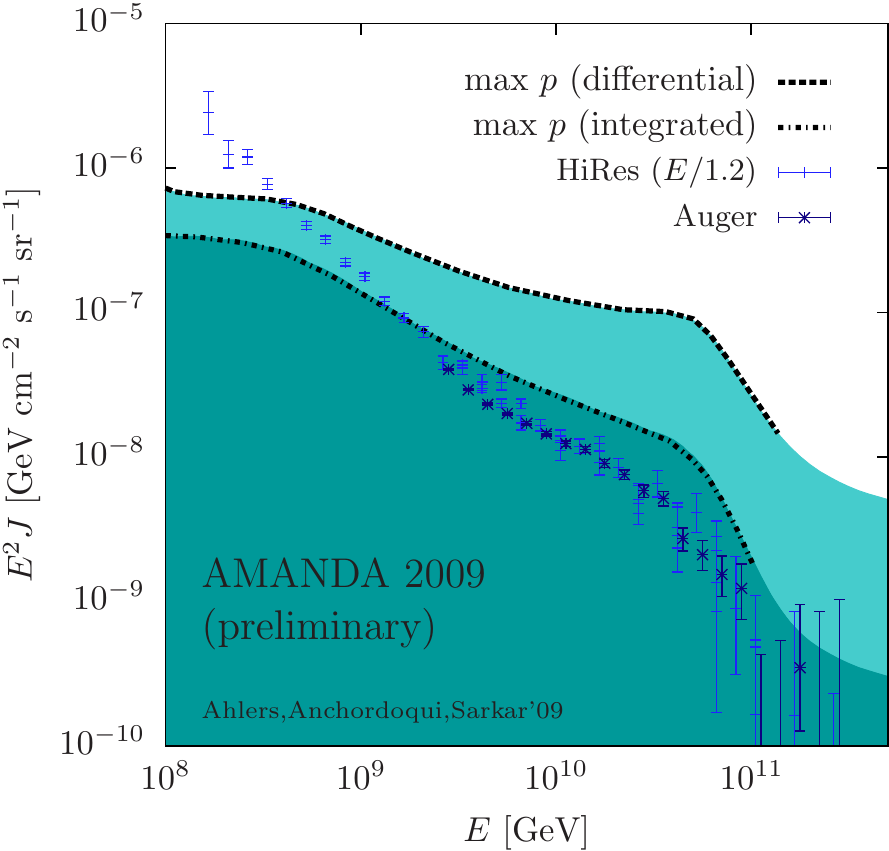}}}
\caption{Integrated and differential upper limits on the proton
  contribution in ultrahigh energy cosmic rays derived from AMANDA
  bound on diffuse neutrinos~\cite{Silvestri}.  (After Ref.~\cite{Ahlers:2009rf}.)}
\label{diagnostics}
\end{figure}

For this procedure one introduces test functions of the neutron source
luminosity of the form $\mathcal{L}^{\rm test}_n(0,E) = \mathcal{L}_0
\, (E/E_\mathrm{max})^{-1}\, \exp (-E/E_\mathrm{max}),$ with an
exponential energy cut-off $E_\mathrm{max}$ that varies between
$10^{8}$~GeV and $10^{12}$~GeV with a logarithmic step-size of $\Delta
\log_{10}E = 0.25$~\cite{Ahlers:2009rf}. The source function per unit
redshift is given in Eq.~(\ref{morton}). The source luminosity per
co-moving volume is assumed to follow that of the star formation rate
(SFR): $\mathcal{H}_{\rm SFR} (z) = (1+z)^{3.4}$, for $z<1$,
$\mathcal{H}_{\rm SFR} (z) = N_1 (1+z)^{-0.3}$, for $1 < z < 4$, and
$\mathcal{H}_{\rm SFR} (z) = N_1 N_4 (1 + z)^{-3.5}$, for $z >4$, with
appropriate normalization factors, $N_1 = 2^{3.7}$ and $N_4 =
5^{3.2}$~\cite{Hopkins:2006bw}. Each neutron test luminosity is
related to a neutrino luminosity by the ratio~(\ref{ratio}). After
propagation, the accumulated contribution of extra-galactic and
cosmogenic neutrinos is normalized to the neutrino limit. The results
are shown in Fig.~\ref{diagnostics}. The AMANDA preliminary limit on
the diffuse neutrino flux~\cite{Silvestri} severely constrains the
extragalactic proton contribution in cosmic rays at energies below
$10^9$~GeV. Following the same line of thought, one can verify that if
the evolution of cosmic ray sources follows that of active galactic
nuclei~\cite{Hasinger:2005sb}, the upper limit on the diffuse flux of
tau neutrinos from the Pierre Auger Observatory~\cite{Abraham:2009eh}
marginally constrains the proton fraction at the end of the
spectrum~\cite{Anchordoqui:2009ty}.

\subsection{Dark matter searches}
\label{S_dark-matter}

Over the past few years, a flood of high-quality
data~\cite{Riess:1998cb} from the Supernova Cosmology Project, the
Supernova Search Team, the Wilkinson Microwave Anisotropy Probe
(WMAP), the Two Degree Field Galaxy Redshift Survey (2dFGRS), and the
Sloan Digital Sky Survey (SDSS) pin down cosmological parameters
to percent level precision, establishing a new paradigm of cosmology.
A surprisingly good fit to the data is provided by a simple
geometrically flat Friedman-Robertson-Walker universe, in which 30\%
of the energy density is in the form of non-relativistic matter
$(\Omega_{\rm m} = 0.30 \pm 0.04$) and 70\% in the form of a new,
unknown dark energy component with strongly negative pressure
$(\Omega_\Lambda = 0.70 \pm 0.04)$. The matter budget has only 3 free
parameters: the Hubble parameter $h = 0.70^{+ 0.04}_{-0.03},$ the
matter density $ \Omega_{\rm m} h^2 = 0.138 \pm 0.012,$ and the
density in baryons, $ \Omega_{\rm b} h^2 =
0.0230^{+0.0013}_{-0.0012}.$\footnote{The latter is consistent with
  the estimate from Big Bang nucleosynthesis, based on measurements of
  deuterium in high redshift absorption systems, $\Omega_{\rm b} h^2 =
  0.020 \pm 0.002$~\cite{Burles:2000zk}.} This implies that the
structure of the universe is dictated by the physics of
as-yet-undiscovered cold dark matter ($\Omega_{\rm CDM} h^2 = 0.115
\pm 0.012$) and the galaxies we see today are the remnants of
relatively small overdensities in the nearly uniform distribution of
matter in the very early universe.

The simplest model for cold dark matter consists of WIMPs - weakly
interacting massive particles~\cite{Bergstrom:2000pn}. The most
popular CDM candidates are {\it (i)} the lightest supersymmetric
particle (such as the well motivated neutralino $\chi$) that is
stabilized in $R$-parity conserving models of supersymmetry
(SUSY)~\cite{Goldberg:1983nd} and {\it (ii)} the lightest Kaluza-Klein
state appearing in models of universal
extra dimensions~\cite{Servant:2002aq}.  Generic WIMPs were once in
thermal equilibrium, but decoupled while strongly
non-relativistic. The relic abundance of SUSY WIMPs can be found by
integrating the Boltzman equation~\cite{Scherrer:1985zt},
\begin{equation}
  \frac{dn}{dt} + 3 H\, n = - \langle \sigma v\rangle (n^2 - n_{\rm{eq}}^2),
\label{expansion}
\end{equation}
where $n$ is the present number density of SUSY WIMPs, $H$ is the
expansion rate of the universe at temperature $T$, $n_{\rm{eq}} = g\,
(m_{\chi}\, T/2 \pi)^{3/2}\,e^{-m_{\chi}/T}$ is the equilibrium number
density, $g$ is the number of internal degrees of freedom of the WIMP,
$\langle \sigma v \rangle$ is the thermally averaged
annihilation cross section, and $m_{\chi}$ is the neutralino
mass. Note that in the very early universe, when $n \simeq
n_{\rm{eq}}$, the right hand side of Eq.~(\ref{expansion}) is small
and the evolution of the density is dominated by Hubble expansion.  As
the temperature falls below $m_{\chi}$, however, the equilibrium
number density becomes suppressed and the annihilation rate increases,
rapidly reducing the number density.  Finally, when the number density
falls enough, the rate of depletion due to expansion becomes greater
than the annihilation rate, $H \geq n \langle  \sigma v
\rangle$, and the neutralinos {\it freeze out} of thermal
equilibrium. The freeze-out temperature $T_{\rm f}$ depends
logarithmically upon $\langle  \sigma v \rangle,$ but for
models with TeV scale SUSY breaking, one finds that $T_{\rm
  f}/m_{\chi} \sim 0.05$. SUSY WIMPs are, by far, the favored
candidate for CDM, because for masses of order 100~GeV to
10~TeV a present density of $\Omega_{\chi}\ h^2 \sim 0.1,$
comes out fairly natural~\cite{Griest:1989zh}.

Many approaches have been developed to attempt to detect dark
matter. Such endeavors include direct detection experiments which hope
to observe the scattering of dark matter particles with the target
material of the detector~\cite{Goodman:1984dc} and indirect detection
experiments which are designed to search for the products of WIMP
annihilation into gamma-rays, anti-matter, and neutrinos~\cite{Silk:1985ax}.

The annihilation of WIMPs into photons typically proceeds through a
complex set of processes, and the final spectrum actually contains a
detailed imprint of WIMP annihilation that can, in principle, reveal
features such as the WIMP spin and/or other particles in the dark
sector~\cite{Bergstrom:1997fj}. On the one hand, tree-level
annihilation of WIMPs into quarks or leptons (or heavier states which
decay into them) yields a continuum of photon energies, with an upper
cutoff at approximately the WIMP mass. On the other hand, loop-level
annihilation into a photon and X results in a line at energy $E_\gamma
= m_\chi [1-M_X^2/(4m_\chi^2)],$ where $M_X$ is the $X$-boson in the
final state (in SUSY models, $X$ is either another photon or a
$Z$). The line emission typically has smaller magnitude than continuum
emission, but it provides a clean signal that helps discriminate
against backgrounds.  The differential flux of photons arising from
CDM annihilation observed in a given direction making an angle $\psi$
with the direction of the galactic center (GC) is given
by~\cite{Bergstrom:1997fj}
\begin{equation}
\phi^{\gamma} (\psi, E_\gamma) = \int \bar{J} \  \frac{1}{2} \ \frac{D_{\odot}}{4\pi} \ \frac{\rho^2_{\odot}}{m_\chi^2} \ \sum_f \, \langle \sigma v \rangle_f \ \frac{dN_f}{dE_{\gamma}} \, d\Omega,
\label{flux} 
\end{equation}
where $\bar{J} = (1/\Delta \Omega) \int_{\Delta \Omega}
J(\psi)\,d\Omega$ denotes the average of $J$ over the solid angle
$\Delta \Omega$ (corresponding to the angular resolution of the
instrument) normalized to the local density: $J(\psi) =
(D_\odot\rho^2_{\odot})^{-1} \int_{\ell =0}^\infty \rho^2[r(\ell,
\psi)] d\ell$; the coordinate $\ell$ runs along the line of sight,
which in turn makes an angle $\psi$ with respect to the direction of
the GC ( i.e., $r^2=\ell^2+D^2_\odot-2 \ell D_\odot \cos{\psi}$); the
subindex $f$ denotes the annihilation channels with one or more
photons in the final state and $dN_f/ dE_\gamma$ is the (normalized)
photon spectrum per annihilation; and $\rho(\vec x)$, $\rho_\odot =
0.3~{\rm GeV}/{\rm cm}^3$, and $D_\odot \simeq 8.5~{\rm kpc}$
respectively denote the CDM density at a generic location $\vec x$
with respect to the GC, its value at the solar system location, and
the distance of the Sun from the GC. Indeed, the GC has long been
considered to be among the most promising targets for detection of CDM
annihilation, particularly if the halo profile of the Milky Way is
cusped in its inner volume~\cite{Navarro:1996gj}. This has been
complicated, however, because H.E.S.S. observations disfavored CDM
emission from the GC, and revealed instead a VHE source (HESS
J1745-290)~\cite{Aharonian:2006wh}. To discover CDM photon emission
the contribution of the many point sources near
GC~\cite{Aharonian:2006au} needs to be subtracted.  IACTs or satellite
and ballon-borne experiments also look for gammas from high CDM
density regions relatively close to the Earth ($\gtrsim 100$ kpc) such
as dwarf spheroidal satellite galaxies of the Milky Way, clusters of
galaxies, and intermediate mass black holes~\cite{Montaruli:2009rr}.

Neutrino telescopes indirectly search for the presence of dark matter
by taking advantage of the Sun's ability to capture large numbers of
WIMPs over time. Over billions of years, a sufficiently large number
of WIMPs can accumulate in the Sun's core to allow for their
efficient annihilation. Such annihilations produce a wide range of
particles, most of which are quickly absorbed into the solar
medium. Neutrinos, on the other hand, may escape the Sun and be
detected in experiments on the Earth. Specifically, the time evolution
of the WIMP population in the Sun is controlled by
\begin{equation}
\dot N = C_\odot - A_\odot N^2 \,,
\label{dm_anni}
\end{equation}
where 
\begin{equation}
  C_\odot \approx 3 \times 10^{18}~{\rm s^{-1}} \, 
\left(\frac{\rho_{\rm local}}{0.3~{\rm GeV}/{\rm cm}^{3}} \right) 
\left(\frac{270~{\rm km}/s}{\overline v_{\rm local}} \right)^3 
\left(\frac{\sigma_{\rm eff}}{10^{-6}~{\rm pb}} \right)  
  \left(\frac{1~{\rm TeV}}{m_\chi} \right)^2
\end{equation}
is the Sun's rate of capture~\cite{Gould}, $\rho_{\rm local}$ is the
local dark matter density, $\overline v_{\rm local}$ is the local rms
velocity of halo dark matter particles, $A_\odot = \langle 
\sigma v \rangle / V_{\rm eff}$, $V_{\rm eff} = 1.8 \times 10^{26}
({\rm TeV}/m_{\chi})^{3/2}~{\rm cm^3}$ is the effective volume of the
core Sun determined by matching the core temperature with the
gravitational potential energy of a single WIMP at the core
radius~\cite{Griest:1986yu}, and $A_\odot$ is related to the
annihilation rate by
\begin{equation}
\Gamma = \frac{1}{2} A_\odot  N^2 \, .
\end{equation}
The rate at which WIMPs are captured in the Sun depends on the nature
of the interaction the WIMP undergoes with nucleons in the Sun. These
elastic scattering processes are often broken into two
classifications: {\it spin dependent} interactions in which cross
sections, $\sigma_{\rm SD},$ increase with the spin of the target
nuclei, and {\it spin independent} interactions in which cross
sections, $\sigma_{\rm SI}$, increase as the square of the total number of
nucleons in the target. Thus, the effective WIMP-on-nucleus (hydrogen
and helium) elastic scattering cross section has two contributions:
$\sigma_{\rm eff} = \sigma_{\rm SD} + \sigma_{\rm ID}.$ The present
WIMP annihilation rate is found by solving Eq.~(\ref{dm_anni}) for the
annihilation rate at a given time
\begin{equation}
\Gamma = \frac{1}{2} C_\odot \, {\rm tanh}^2 
\left(\frac{t}{\tau_{\rm eq}}\right)\, ,
\end{equation}
where $\tau_{\rm eq} = 1/\sqrt{C_\odot \, A_\odot}$ is the time scale
required to reach equilibrium between capture and annihilation. When
$\tau_{\rm eq}$ becomes comparable or larger than the age of the solar
system ($t_\odot \simeq 4.5$ billion years), the system has not yet
reached equilibrium and the annihilation rate is strongly suppressed
$\Gamma \approx \frac{1}{2} \, C_\odot (\sqrt{C_\odot \, A_\odot} \,
t_\odot)^2$, whereas when $(C_\odot \, A_\odot) \, t_\odot \gg 1$ the
annihilation rate is saturated at $\Gamma = \frac{1}{2} C_\odot$.

As they annihilate, WIMPs can generate neutrinos through a wide range
of channels. Annihilations to heavy quarks, tau leptons, gauge bosons
and Higgs bosons can all generate neutrinos in the subsequent 
decay~\cite{Jungman:1994jr}.
The total flux of neutrinos emitted by the Sun due to WIMP 
annihilition is then
\begin{equation} 
\left. \frac{dF^{\nu_\mu}_\odot}{dE_\nu}\right|_{\rm source} = C_\odot \, F_{\rm Eq} \sum_j 
\left(\frac{dN_{\nu_\mu}}{dE_\nu}\right)_{\! j}\,\,\,  
e^{-E_\nu/150~{\rm GeV}}\, \, ,
\end{equation}
where $F_{\rm{Eq}}$ is the non-equilibrium suppression factor
($\approx 1$ for capture-annihilation equilibrium) and
$(dN_{\nu_\mu}/dE_\nu)_j$ is the $\nu_\mu$ flux produced by the $j$
decay channel per WIMP annihilation.  Note that neutrinos produced
near the center of the Sun interact with the solar medium, yielding a
depletion of the emission spectrum by a factor $\sim
e^{-E_\nu/150~{\rm GeV}}$~\cite{Jungman:1994jr}. Finally, the
$\nu_\mu$ flux reaching the Earth is,
\begin{equation}
\phi^{\nu_\mu}_\odot = \frac{1}{4 \pi d^2}
\left. \frac{dF^{\nu_\mu}_\odot}{dE_\nu}\right|_{\rm source} \,\,,
\end{equation}
where $d \approx 1.5 \times 10^8~{\rm km}$ is the Earth-Sun distance.
All in all, the prospects for neutrino-detection-experiments detecting
dark matter critically depend on the capture rate of WIMPs in the
Sun, which in turn depends on the elastic scattering cross section of
these particles.

\begin{figure}[tpb]
\rotatebox{0}{\resizebox{12.5cm}{!}{\includegraphics{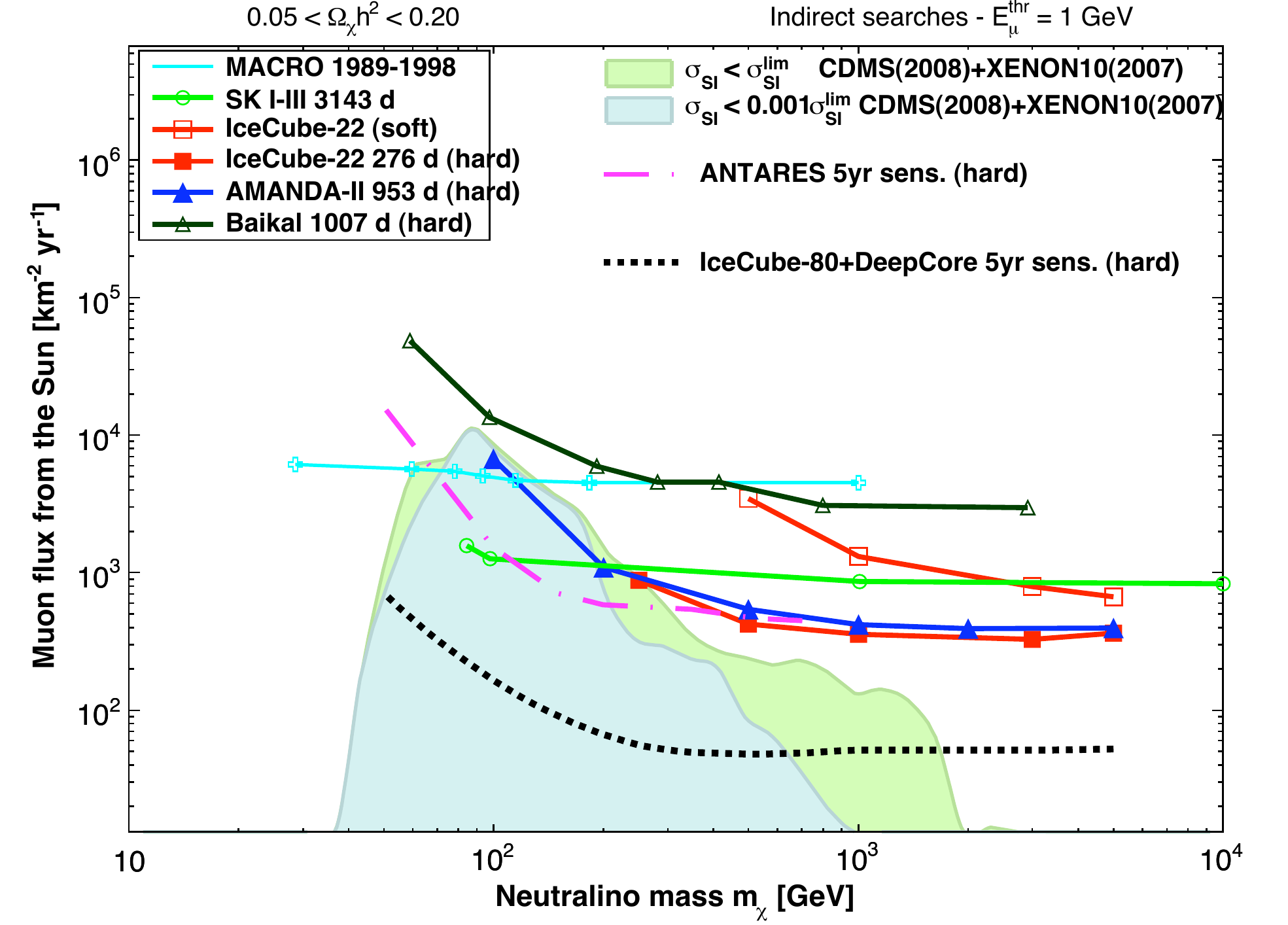}}}
\caption{Upper limits (90\%CL) on the muon fluxes from neutralino
  annihilation in the Sun as a function of neutralino mass for soft
  ($b\bar{b}$) and hard ($W^+W^-$) channels~\cite{Abbasi:2009uz,Montaruli:2009rr}. The lighter hatched
  region indicates the SUSY parameter space compatible with direct
  detection limits on the spin independent cross section from
  CDMS~\cite{Ahmed:2008eu} and XENON10~\cite{Angle:2007uj}. The darker
  hatched area represents the projected sensitivity for the same
  region of the parameter space, assuming that direct detection limits
  are a 100 times better.  Experimental limits, that include a
  correction for detector threshold within the common assumption of
  $E_{\nu, {\rm thr}} = 1~{\rm GeV}$, are shown for
  Super-Kamiokande~\cite{dm_sk}, MACRO~\cite{Ambrosio:1998qj},
  AMANDA-II~\cite{dm_amanda}, Baikal~\cite{dm_baikal} and IceCube 22
  strings for the hard and soft channels~\cite{Abbasi:2009uz}. For
  comparison, the projected sensitivity of ANTARES after 5~yr of data
  taken is also shown \cite{dm_antares}.} 
\label{dmfig1}
\end{figure}

\begin{figure}
\rotatebox{0}{\resizebox{12.5cm}{!}{\includegraphics{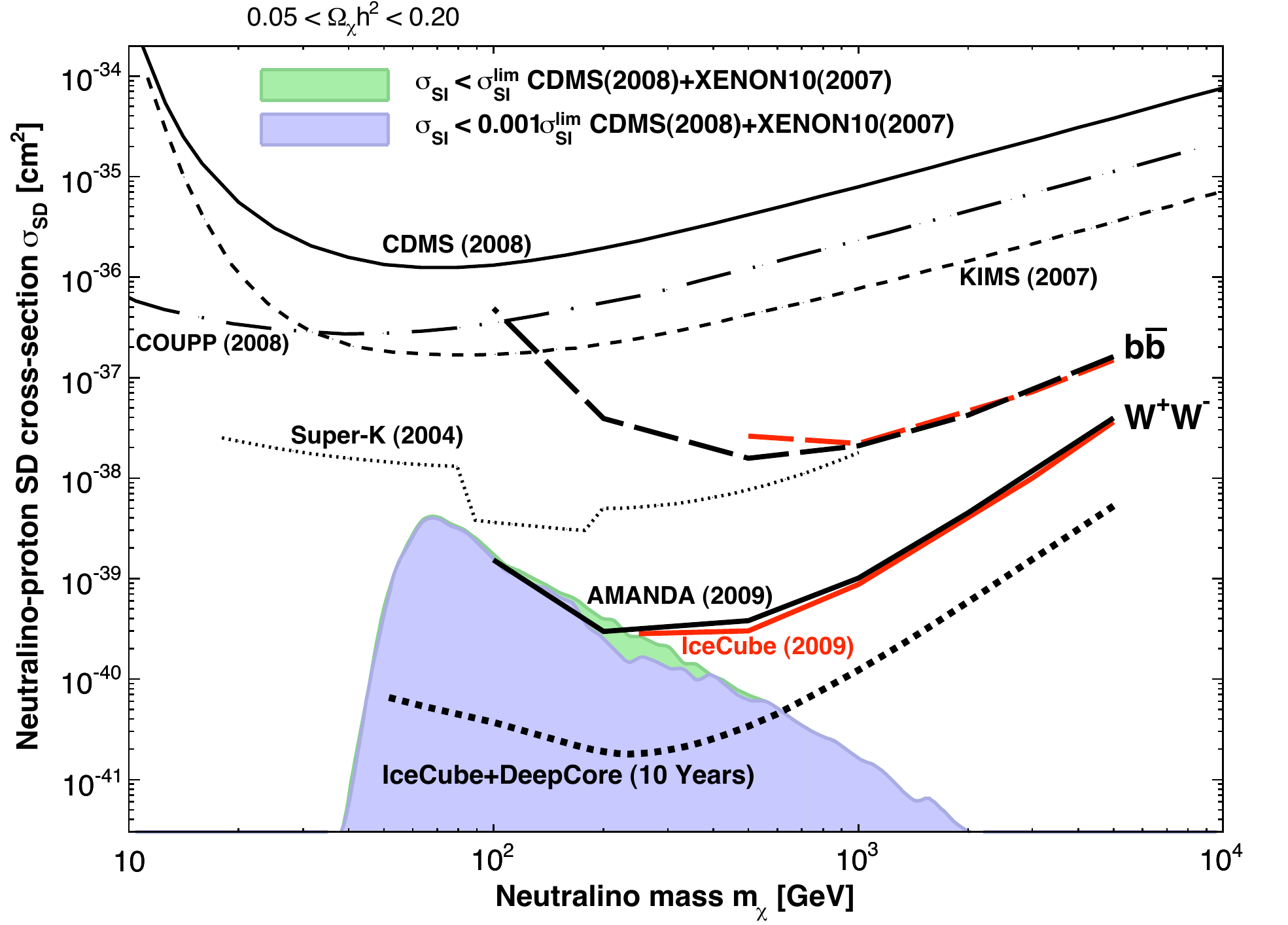}}}
\caption{90\% CL upper limits from
  AMANDA-II and 22 strings of IceCube on the spin dependent cross
  section for soft and hard channels~\cite{Abbasi:2009uz}. Hatched
  regions follow the same conventions of Fig.~\ref{dmfig1}. For
  comparison, we also show direct search limits on $\sigma_{\rm SD}$
  from CDMS~\cite{Ahmed:2008eu}, COUPP~\cite{Behnke:2008zza} and
  KIMS.
  }
\label{dmfig2}
\end{figure}

While gamma-ray experiments set limits on $\langle  \sigma v
\rangle$ or on the gamma-ray flux, neutrino telescopes measure the muon
induced flux by neutrinos and hence constrain this flux as a function
of $m_\chi$. A collection of various limits on the muon flux is shown
in Fig.~\ref{dmfig1}. The conversion from muon flux to cross section
limits is not model independent; in Fig.~\ref{dmfig2} it was done using
DarkSUSY~\cite{Gondolo:2004sc}. Equilibrium is assumed between capture
and annihilation rates in the Sun, so that the annihilation rate is
proportional to the spin-dependent and independent cross sections. A
limit on $\sigma_{\rm SD}$ is found by setting to zero the
spin-independent cross section. This procedure is indispensable to
show complementarity between searches and to combine results from
different techniques. As an illustration, in Fig.~\ref{dmfig2} we
compare the limits on $\sigma_{\rm SD}$ with those obtained from
direct detection experiments including CDMS~\cite{Ahmed:2008eu},
COUPP and KIMS~~\cite{Behnke:2008zza}.

The described indirect searches for CDM are complementary approaches
to the direct searches and the possible direct production at
LHC. Though indirect detection channels are subject to large astrophysical
uncertainties, they explore different regions of the parameter space
and a very wide range of candidate particle masses.

\section{CONCLUSIONS AND FUTURE PROSPECTS}
\label{sec5}

The hunt for neutrinos has just begun with suitable size detectors for
the expected small event rates. Given their dimensions these
experiments are very challenging to build. IceCube, the first
cubic-kilometer detector, is expected to be completed on schedule in
2011.  If the predictions from the measured gamma-ray fluxes are
indeed correct, then the first detection of sources, such as the
``Milagro Pevatrons'' with hard spectra up to a few tens of TeV, may
become possible after a few years of data taking.  Neutrino telescopes
may broaden our reach to the unknown like many other astronomical
instruments which have discovered unexpected signals from the
Universe; NTs can access sources that are opaque in photons.  The
Mediterranean community is operating ANTARES, the first complete
underwater detector, at the scale of the IceCube's precursor
AMANDA-II.  Other R\&D programs are ongoing to build a cubic-km
detector in the Mediterranean sensitive to the GC. In Italy, a $\sim
100$~km electro-optical cable has been installed off-shore Capo
Passero (at about 3.5 km below sea surface) by the NEMO Collaboration,
in July 2007.  A prototype tower of the proposed cubic-km array will
be connected in the near future. ANTARES, NEMO and NESTOR joined their
efforts forming the Consortium KM3NeT to design the cubic-km detector
in the Mediterranean. High quantum efficiency PMTs are being studied
for this cubic-km detector in the Mediterranean.  Most of the
proposals aim at increasing the photocathode area per OM and the
isotropy of light detection, minimizing the transit time spread,
maximizing the collection efficiency and the peak to valley ratio for
a single photoelectron signal (connected to the capability of counting
photoelectrons in one event), and minimizing costs and power
consumption.

Building a detector a factor of 10 more sensitive than IceCube does not seem
to be an easy task to achieve, neither in the ice nor in the sea. A
new detector in the sea, with similar cost to IceCube, could not
achieve such a goal with the current technology, despite the fact that
in the sea the angular resolution for point source searches can be
better by a factor of 2-3 compared to IceCube. It is also hard  to
obtain a uniform increment in area at all energies, while avoiding cost divergence. A
significant increase in the area at very large energies implies larger
separations between instrumented lines. Hence, this penalizes the
1$-$100~TeV region, that is an important region for galactic sources.

Projects are underway to build detectors of the order of 100 times
IceCube using techniques that exploit the advantage of attenuation
lengths about a factor of 10 greater than the optical one in the same
medium. Particularly interesting is the radio technique in ice. The
acoustic technique in sea water or ice seems also encouraging.  None
of these techniques, however, is background free. Moreover, they have
the disadvantage of higher energy thresholds compared to the NTs
discussed in this review. Clearly the higher thresholds reduces the
degree of overlap with the optical technique, preventing effective
cross-calibrations that would help the understanding of the background.  Moreover, thresholds of the order of $10^{18}$~eV
are suitable for extragalactic neutrinos but not for galactic sources.
Detectors combining the radio and acoustic techniques are also
desirable. However, for cross-calibrations, the radio technique seems
currently more promising due to the low cost of antennas, the lower
power consumption than acoustic devices, and the larger attenuation
length in the ice.

In summary, after 20~years of careful work by many research groups around
the world we are in possession of a tantalizing body of data, more
than sufficient to stimulate our curiosity but not yet sufficient to
pin down high energy neutrino sources. The upcoming high quality
observations from the ``giant-aperture'' neutrino telescopes under
construction will generate a data sample of unprecedented size,
ushering us to a golden age of neutrino astronomy.

\section*{Acknowledgments}

We are grateful to Markus Ahlers, Francis Halzen, Luciano Moscoso, Paolo Desiati, Ellen Zweibel, Christian Spiering, Tom Gaisser, Todor Stanev, Tom Weiler, Francesco Arneodo, Haim Goldberg,  Subir Sarkar, Diego Torres, Evelyn Malkus for comments and discussions. L.A.A.\ is supported by the U.S. National Science Foundation Grant No PHY-0757598, and the UWM Research Growth Initiative.  T.M.\  is supported by U.S. National Science Foundation-Office of Polar Program, U.S. National Science Foundation-Physics Division, and University of Wisconsin Alumni Research Foundation and by INFN, Sezione di Bari, Italy.

\end{document}